\documentclass[10pt,twocolumn]{article}

\usepackage[margin=.9in]{geometry}
\usepackage{natbib}
\usepackage{arydshln}
\usepackage{booktabs}
\usepackage{enumitem}
\usepackage{amssymb,mathtools}
\usepackage[colorlinks,allcolors=blue]{hyperref}
\usepackage[small]{titlesec}
\usepackage{amsmath,amsthm}
\usepackage{pgfplots}
\usepackage[flushleft]{threeparttable}
\usepackage[bf,normalsize]{caption}
\usepackage{multirow}
\usepackage{setspace}
\usepackage{anyfontsize}
\usepackage{comment}
\usepackage{subcaption}
\usepackage[capposition=top]{floatrow}
\usepackage{stfloats}

\theoremstyle{definition}
\newtheorem*{LP*}{LP*}

\newtheorem*{LP}{LP}
\newtheorem*{MILP}{MILP}
\newtheorem*{MILP-ST}{MILP-ST}

	\author{{\normalsize Brendan K.\ Beare} \\ {\normalsize School of Economics} \\ {\normalsize University of Sydney} \\ {\normalsize \href{mailto:brendan.beare@sydney.edu.au}{brendan.beare@sydney.edu.au}} \and {\normalsize Juwon Seo} \\ {\normalsize Department of Economics} \\ {\normalsize National University of Singapore} \\ {\normalsize \href{mailto:ecssj@nus.edu.sg}{ecssj@nus.edu.sg}} \and {\normalsize Zhongxi Zheng} \\ {\normalsize Department of Economics} \\ {\normalsize National University of Singapore} \\ {\normalsize \href{mailto:zhongxi.zheng@u.nus.edu}{zhongxi.zheng@u.nus.edu}}}

\begin{document}

	\title{Stochastic arbitrage with market index options}
	
	\twocolumn[
	\begin{@twocolumnfalse}
	\maketitle
	\center{Accepted for publication in the \emph{Journal of Banking \& Finance}.}
	\bigskip
	\begin{abstract}
		Opportunities for stochastic arbitrage in an options market arise when it is possible to construct a portfolio of options which provides a positive option premium and which, when combined with a direct investment in the underlying asset, generates a payoff which stochastically dominates the payoff from the direct investment in the underlying asset. We provide linear and mixed-integer linear programs for computing the stochastic arbitrage opportunity providing the maximum option premium to an investor. We apply our programs to 18 years of data on monthly put and call options on the Standard \& Poors 500 index, finding no evidence that stochastic arbitrage opportunities are systematically present. A skewed specification of the underlying market return distribution with a constant market risk premium and constant multiplicative variance risk premium is broadly consistent with the pricing of market index options at moderate strikes. 
		\bigskip
	\end{abstract}
\end{@twocolumnfalse}
	]
\section{Introduction}\label{sec:intro}
This article concerns the possibility of engaging in stochastic arbitrage using market index options. To set the scene, consider a simple one-period model in which there is a single risky underlying asset, to be thought of as a market index, and a collection of options (derivative securities) written on the underlying asset. The payoff of each option after one period is determined by the value of the underlying asset. An investor takes a unit position in the underlying asset and considers augmenting this position with a portfolio of options called a \emph{layover portfolio}, which may in general contain both long and short positions in options. We say that a layover portfolio is a \emph{first-order stochastic arbitrage opportunity} if (i) it provides a positive option premium, i.e.\ its price is negative, and (ii) the payoff distribution of the layover portfolio combined with a unit investment in the underlying asset first-order stochastically dominates the payoff distribution of the unit investment in the underlying asset. We define a \emph{second-order stochastic arbitrage opportunity} in the same way, but with second-order stochastic dominance replacing first-order stochastic dominance.

An algorithm for computing the second-order stochastic arbitrage opportunity with maximum premium whenever such an opportunity exists has recently been provided in \citet{PostLongarela21}. The algorithm is a linear program. The objective function to be maximized is the negative of the price of a candidate layover portfolio, with bid and ask prices applied to short and long positions in different options. The constraints in the linear program require positions in options to be bounded by the quantities available for sale or purchase at each bid or ask price, and require the payoff distribution of the layover portfolio combined with a unit position in the underlying asset to second-order stochastically dominate the unit position in the underlying asset. If the maximum value of the objective function is positive, then the corresponding layover portfolio is the maximum premium second-order stochastic arbitrage opportunity. If the maximum value of the objective function is zero (which is always feasible with zero portfolio weights) then no second-order stochastic arbitrage opportunity exists. The linear program developed in \citet{PostLongarela21} builds upon a long literature dealing with stochastic dominance pricing bounds for options, with fundamental earlier contributions including \citet{PerrakisRyan84}, \citet{Levy85}, \citet{Ritchken85} and \citet{ConstantinidesJackwerthPerrakis09}; see \citet{Perrakis19} for further discussion and references.

We make two primary methodological contributions concerning the computation of stochastic arbitrage opportunities. First, we show that the linear program developed in \citet{PostLongarela21} may be reformulated in a way which reduces the number of inequalities and equalities by an order of magnitude, leading to a concomitant reduction in runtime. This is achieved using an efficient linear formulation of second-order stochastic dominance given in \citet{Luedtke08}. Second, we provide a mixed-integer linear program for computing the first-order stochastic arbitrage opportunity with maximum premium. This is achieved using a formulation of first-order stochastic dominance given in \citet{Luedtke08} in which linear inequality and equality constraints are combined with binary constraints on certain auxiliary choice variables.

We report the results of a substantive empirical application to monthly Standard \& Poors 500 index (SPX) options. This options market is a particular focus of the extensive literature on the so-called pricing kernel puzzle originating with \citet{AitSahaliaLo00}, \citet{Jackwerth00} and \citet{RosenbergEngle02}. In each of these articles an empirical estimate of the pricing kernel (ratio of Arrow security prices to state probabilities) implied by vanilla SPX options was found to be locally increasing around the center of the SPX return distribution, contradicting the nonincreasing property of market index pricing kernels predicted by standard financial theory. More recent articles reporting similar evidence include \citet{BakshiMadanPanayotov10}, \citet{ChaudhuriSchroder15}, \citet{BeareSchmidt16} and \citet{CuesdeanuJackwerth18}. On the other hand, articles including \citet{ChabiYoGarciaRenault08}, \citet{ChristoffersenHestonJacobs13}, \citet{SongXiu16} and \citet{LinnShiveShumway18} argue that findings of pricing kernel nonmonotonicity may be a spurious outcome of failing to properly condition on relevant information in the construction of state probabilities.

There is a fundamental connection between the presence of locally increasing regions in a pricing kernel and the presence of stochastic arbitrage opportunities. It is shown in \citet{Dybvig88} and \citet{Beare11} that, in a stylized model with a complete and frictionless options market, pricing kernel nonmonotonicity is equivalent to the existence of a first-order stochastic arbitrage opportunity. Relatedly, it is shown in \citet{PostLongarela21} that, in a model permitting bid-ask spreads and an incomplete options market, the nonexistence of a monotone pricing kernel respecting bid-ask spreads is equivalent to the existence of a second-order stochastic arbitrage opportunity. Closely related theoretical results concerning the existence of first- and second-order stochastic arbitrage opportunities are provided in \cite{Beare23}. See \citet{Perrakis22} for further discussion of the link between pricing kernel monotonicity and the existence of stochastic arbitrage opportunities.

An indirect test for pricing kernel nonmonotonicity---i.e., the pricing kernel puzzle---may be undertaken by directly assessing whether there exist stochastic arbitrage opportunities in the market for vanilla SPX options. In our empirical application, for one day in each month from January 2004 to November 2021, we use our linear and mixed-integer linear programs to search for second- and first-order stochastic arbitrage opportunities in SPX options with one month to expiry. These programs require a specification of the SPX return distribution as an input. Our specification uses a skewed generalized t-distribution centered at the risk-free rate plus a constant market risk premium and scaled by the VIX deflated by a constant multiplicative variance risk premium. We report separate analyses for strikes ranging from 10\% below to 5\% above the current SPX value, and for strikes ranging from 30\% below to 15\% above the current SPX value. We refer to these two strike ranges as \emph{moderate} and \emph{wide}.

It may be useful to consider three possible outcomes to a search for stochastic arbitrage opportunities across numerous months using a particular specification of the SPX return distribution.
\begin{enumerate}
	\item Stochastic arbitrage opportunities are rarely identified.\label{en:Outcome1}
	\item Stochastic arbitrage opportunities are identified in a substantial fraction of months, but deliver realized losses.\label{en:Outcome2}
	\item Stochastic arbitrage opportunities are identified in a substantial fraction of months, and deliver realized profits.\label{en:Outcome3}
\end{enumerate}
Outcome \ref{en:Outcome1} indicates that the pricing of options is broadly consistent with the specified SPX return distribution and, if that specification is viewed as plausible, provides evidence against the presence of economically meaningful pricing kernel nonmonotonicity. Outcome \ref{en:Outcome2} indicates an undesirable specification of the SPX return distribution. Outcome \ref{en:Outcome3} provides evidence that the pricing kernel puzzle is real: options are fundamentally mispriced. In different parts of our empirical analysis we observe Outcomes \ref{en:Outcome1} and \ref{en:Outcome2}, but never Outcome \ref{en:Outcome3}.

Outcome \ref{en:Outcome1} is obtained when we search for stochastic arbitrage opportunities within the moderate strike range using the skewed generalized t-distribution to specify market return probabilities. Substantial stochastic arbitrage opportunities are identified in only a small minority of months, and do not generate superior realized returns. This may be surprising in view of earlier literature on the pricing kernel puzzle discussed above, because if option prices indicate that the pricing kernel is locally increasing near the center of the SPX return distribution then it ought to be possible to identify profitable stochastic arbitrage opportunities within our moderate strike range. We provide evidence that the pricing kernel nonmontonicity identified in \citet{Jackwerth00}, \citet{RosenbergEngle02} and elsewhere may be due in part to insufficient asymmetry near the center of the specified SPX return distribution.

Our empirical results are somewhat different when we extend our search for stochastic arbitrage opportunities to the wider strike range spanning from 30\% below to 15\% above the current SPX. In this range of strikes we identify a substantial number of second-order, but not first-order, stochastic arbitrage opportunities. The realized performance of the identified second-order stochastic arbitrage opportunities is mixed, but overall worse than simply holding the market portfolio when pure arbitrage opportunities are excluded. The stochastic arbitrage opportunities identified generally involve taking short positions in far out-of-the-money call options, thereby generating superior returns in months where the SPX return is not very large but potentially suffering large losses when the SPX performs very well. Several such large losses occurring in the latter half of our sample period drag down the overall performance of the selected option portfolios. Thus we have Outcome \ref{en:Outcome2}, which is perhaps unsurprising given the inherent difficulty of accurately specifying probabilities far in the tails of the SPX return distribution.

The remainder of this article is structured as follows. In Section \ref{sec:problem} we define the option portfolio choice problem to be solved in each period. Our linear and mixed-integer linear programs for solving the portfolio choice problem are described in Section \ref{sec:sd}. Section \ref{sec:empirical} presents the results of our empirical implementation with SPX options. We offer some concluding remarks in Section \ref{sec:conclusion}. In Appendix \ref{sec:LPstar} we demonstrate the superior computational efficiency of the linear program described in Section \ref{sec:sd} relative to the equivalent linear program described in \cite{PostLongarela21}. In Appendix \ref{sec:MILP} we provide a method for choosing starting values for our mixed-integer linear program which greatly reduces the runtime required to obtain a satisfactory solution. In Appendix \ref{sec:tv} we explore the effect of introducing time-varying risk premia or asymmetry to our model of state probabilities, finding that the central conclusions of our article are largely unaffected.

\section{Portfolio choice problem}\label{sec:problem}

We take the perspective of an investor holding one unit of the underlying asset. At a fixed future time (say, one month hence) this unit is sold and delivers the investor a nonnegative random payoff taking values in a finite set of points $x_1<\cdots<x_n$. The probabilities with which the random payoff takes each of these values are denoted $\mu_1,\dots,\mu_n$. There are $m$ options written on the underlying asset, each of which delivers a nonnegative payoff after one month determined by the payoff delivered by the underlying asset. We denote the payoff delivered by one unit of the $i$th of the $m$ options when the underlying asset payoff is $x$ by $\theta_{i}(x)$, and we define $\theta_{ij}=\theta_i(x_j)$. For instance, if the $i$th option is a call option with strike $s_i$, then $\theta_{ij}=\max\{0,x_j-s_i\}$. The investor may take a long position in the $i$th option at price $p_i>0$ per unit, or a short position at price $q_i\geq0$ per unit, where $p_i\geq q_i$.

We use boldface notation for vectors and matrices, and uppercase notation for matrices. Write $\mathbf{\Theta}$ for the $m\times n$ matrix with entries $\theta_{ij}$, write $\boldsymbol{\mu}$ and $\mathbf{x}$ for the $n\times1$ vectors with entries $\mu_j$ and $x_j$ respectively, and write $\mathbf{p}$ and $\mathbf{q}$ for the $m\times1$ vectors with entries $p_i$ and $q_i$ respectively.

A layover portfolio is represented by two $m\times1$ nonnegative vectors $\boldsymbol{\alpha},\boldsymbol{\beta}$ indicating the respective long and short positions taken by the investor in each option. The price of such a portfolio is $\mathbf{p}^\top\boldsymbol{\alpha}-\mathbf{q}^\top\boldsymbol{\beta}$. We call the negative of the price of a layover portfolio its premium. We suppose that long and short positions are constrained to belong to a polytope
\begin{align*}
	\mathcal P&=\{(\boldsymbol{\alpha},\boldsymbol{\beta})\in\mathbb R_+^m\times\mathbb R_+^m:\mathbf{A}\boldsymbol{\alpha}+\mathbf{B}\boldsymbol{\beta}\leq\mathbf{c}\},
\end{align*}
where $\mathbf{A}$ and $\mathbf{B}$ are $\ell\times m$ matrices and $\mathbf{c}$ is an $\ell\times1$ vector, for some $\ell$. For instance, if $\mathbf{v}$ is an $m\times1$ vector whose entries are the maximum long positions that can be taken in each option, and $\mathbf{w}$ is an $m\times1$ vector whose entries are the maximum short positions that can be taken in each option, then we can constrain the investor to respect these limits by setting
\begin{equation}\label{eq:positionlimits}
	\mathbf{A}=\begin{bmatrix}	\mathbf{I}_m\cr\mathbf{0}_{m\times m}\end{bmatrix},\,\mathbf{B}=\begin{bmatrix}\mathbf{0}_{m\times m}\cr\mathbf{I}_m\end{bmatrix},\,\mathbf{c}=\begin{bmatrix}\mathbf{v}\cr\mathbf{w}\end{bmatrix},
\end{equation}
where $\mathbf{I}_m$ is an $m\times m$ identity matrix and $\mathbf{0}_{m\times m}$ is an $m\times m$ zero matrix.

Let $x$ be the random payoff of the underlying risky asset. We suppose that the investor chooses a layover portfolio $(\boldsymbol{\alpha},\boldsymbol{\beta})$ to solve the following optimization problem:
\begin{align*}
	\text{maximize}\quad&-\mathbf{p}^\top\boldsymbol{\alpha}+\mathbf{q}^\top\boldsymbol{\beta}\\
	\text{subject to}\quad&x+\sum_{i=1}^m\alpha_i\theta_i(x)-\sum_{i=1}^m\beta_i\theta_i(x)\gtrsim x,\\
	&(\boldsymbol{\alpha},\boldsymbol{\beta})\in\mathcal P.
\end{align*}
Here, $\gtrsim$ denotes either first- or second-order stochastic dominance. In the latter case our optimization problem is the same as in \citet{PostLongarela21}.

The options to be traded in our empirical application are put and call options. In a departure from previous studies, we require the layover portfolio to deliver zero payoff when the underlying asset falls outside the range of strikes at which options are traded. This constraint may exclude certain stochastic arbitrage opportunities from consideration. From a practical perspective, it may be a reasonable restriction for an investor to impose if they feel unable to reliably specify probabilities for states falling outside the range of strikes, or if they wish to insulate their position in options from extreme movements in the underlying asset.

In order to enforce our requirement that the layover portfolio deliver zero payoff outside the range of strikes, we augment $\mathbf{A}$, $\mathbf{B}$ and $\mathbf{c}$ in \eqref{eq:positionlimits} with 8 additional rows. For $i=1,\dots,m$, let $s_i$ denote the strike of the $i$th option and let $d_i=0$ if the $i$th option is a put or $d_i=1$ if the $i$th option is a call. Order the options such that $s_1\leq\cdots\leq s_m$. (Each strike is repeated once in the empirically common case where both a put and a call are listed at each strike.) Let $\mathbf{a}_{2m+1}$, $\mathbf{a}_{2m+3}$, $\mathbf{a}_{2m+5}$ and $\mathbf{a}_{2m+7}$ be $1\times m$ vectors with $i$th entries equal to $d_is_i$, $d_i$, $(1-d_i)s_i$ and $1-d_i$ respectively. We specify the polytope $\mathcal P$ by setting
\begin{equation}\label{eq:positionlimitsaug}
	\mathbf{A}=\begin{bmatrix}	\mathbf{I}_m\cr\mathbf{0}_{m\times m}\cr\mathbf{a}_{2m+1}\cr-\mathbf{a}_{2m+1}\cr\mathbf{a}_{2m+3}\cr-\mathbf{a}_{2m+3}\cr\mathbf{a}_{2m+5}\cr-\mathbf{a}_{2m+5}\cr\mathbf{a}_{2m+7}\cr-\mathbf{a}_{2m+7}\end{bmatrix},\,\mathbf{B}=\begin{bmatrix}\mathbf{0}_{m\times m}\cr\mathbf{I}_m\cr-\mathbf{a}_{2m+1}\cr\mathbf{a}_{2m+1}\cr-\mathbf{a}_{2m+3}\cr\mathbf{a}_{2m+3}\cr-\mathbf{a}_{2m+5}\cr\mathbf{a}_{2m+5}\cr-\mathbf{a}_{2m+7}\cr\mathbf{a}_{2m+7}\end{bmatrix},\,\mathbf{c}=\begin{bmatrix}\mathbf{v}\cr\mathbf{w}\cr0\cr0\cr0\cr0\cr0\cr0\cr0\cr0\end{bmatrix}.
\end{equation}
The 8 inequality constraints corresponding to the 8 final rows of $\mathbf{A}$, $\mathbf{B}$ and $\mathbf{c}$ consist of 4 negative pairings and may thus be rewritten as 4 equality constraints.

The 4 final rows of $\mathbf{A}$, $\mathbf{B}$ and $\mathbf{c}$ enforce the restriction that the layover portfolio deliver zero payoff below the range of strikes, and the 4 rows immediately above enforce the restriction that the layover portfolio deliver zero payoff above the range of strikes. To see why observe that, since all options are puts or calls, the payoff function of any layover portfolio is continuous and piecewise linear, with kinks permitted only at strikes. It therefore delivers zero payoff below the range of strikes if and only if the payoff function is zero at $s_1$ and has zero left-derivative at $s_1$. This is the case precisely when
\begin{align*}
	\sum_{i=1}^m(1-d_i)(\alpha_i-\beta_i)(s_i-s_1)&=0,\\\sum_{i=1}^m(1-d_i)(\alpha_i-\beta_i)&=0,
\end{align*}
which may be equivalently rewritten as
\begin{equation*}
	\mathbf{a}_{2m+5}\boldsymbol{\alpha}-\mathbf{a}_{2m+5}\boldsymbol{\beta}=0,\quad\mathbf{a}_{2m+7}\boldsymbol{\alpha}-\mathbf{a}_{2m+7}\boldsymbol{\beta}=0.
\end{equation*}
These 2 equalities are equivalent to the 4 inequalities corresponding to the final 4 rows of $\mathbf{A}$, $\mathbf{B}$ and $\mathbf{c}$. A symmetric argument shows that the 4 rows immediately above enforce the requirement that the layover portfolio delivers zero payoff above the range of strikes.

When $\mathbf{A}$, $\mathbf{B}$ and $\mathbf{c}$ are specified to force the layover portfolio to deliver zero payoff outside the range of strikes, as is the case in \eqref{eq:positionlimitsaug}, there is no need to assign probabilities to payoff values that lie outside the range of strikes. This is because, in this case, the first- or second-order stochastic dominance constraint holds \emph{unconditionally} if and only if it holds \emph{conditionally} on $x$ falling within the range of strikes. We therefore need only specify payoff values $x_1,\dots,x_n$ lying within the range of strikes, and assign to them probabilities $\mu_1,\dots,\mu_n$ which are conditional on the random payoff $x$ falling within the range of strikes. Note that conditional state probabilities are obtained simply by scaling the unconditional probabilities of states within the range of strikes such that they sum to one. Large improvements in computational efficiency can be achieved by restricting the payoff values to fall within the range of strikes, because, as we will see in Section \ref{sec:sd}, our optimization problem is solved using a linear or mixed-integer linear program with $n^2+n+2m$ choice variables. We therefore do not want to choose an unnecessarily large number of payoff values. The inclusion of numerous additional values lying outside the range of strikes does not affect the solution to our optimization problem in principle but can lead to it being slow or practically infeasible to compute.

\section{Stochastic dominance constraints}\label{sec:sd}

In order to solve the portfolio choice problem posed in Section \ref{sec:problem} we need to formulate the stochastic dominance constraint in a way that is amenable to available optimization routines. We will discuss second- and first-order stochastic dominance in turn.

\subsection{Second-order stochastic dominance}\label{sec:sd2}

Let $\mathbf{1}_n$ denote the $n\times1$ vector with each entry equal to one, and let $\mathbf{S}$ denote the $n\times n$ strictly lower triangular matrix with all entries below the diagonal equal to one. We adopt a linear formulation of second-order stochastic dominance given in \citet{Luedtke08}. Theorem 4.2 therein, and the preceding discussion, implies that a layover portfolio $(\boldsymbol{\alpha},\boldsymbol{\beta})\in\mathcal P$ satisfies the second-order stochastic dominance constraint
\begin{align}\label{eq:sdconstraint}
	x+\sum_{i=1}^m\alpha_i\theta_i(x)-\sum_{i=1}^m\beta_i\theta_i(x)\gtrsim x
\end{align}
if and only if there exists an $n\times n$ nonnegative matrix $\boldsymbol{\Psi}$ and an $n\times1$ nonnegative vector $\boldsymbol{\xi}$ such that
\begin{align}
	\boldsymbol{\Psi}\mathbf{1}_n&=\mathbf{1}_n,\label{eq:Lconstraint1}\\
	\boldsymbol{\xi}-\boldsymbol{\Psi}^\top\boldsymbol{\mu}&=\mathbf{0}_n,\label{eq:Lconstraint2}\\
	\mathbf{S}\boldsymbol{\xi}&\leq\mathbf{S}\boldsymbol{\mu},\label{eq:Lconstraint3}\\
	\boldsymbol{\Psi}\mathbf{x}-\boldsymbol{\Theta}^\top(\boldsymbol{\alpha}-\boldsymbol{\beta})&\leq\mathbf{x}.\label{eq:Lconstraint4}
\end{align}
It is therefore possible to solve the portfolio choice problem posed in Section \ref{sec:problem} using the following linear program, which we refer to as LP.
\begin{LP}
	Choose two $m\times1$ nonnegative vectors $\boldsymbol{\alpha},\boldsymbol{\beta}$, an $n\times1$ nonnegative vector $\boldsymbol{\xi}$, and an $n\times n$ nonnegative matrix $\boldsymbol{\Psi}$ to maximize $-\mathbf{p}^\top\boldsymbol{\alpha}+\mathbf{q}^\top\boldsymbol{\beta}$ subject to the constraints \eqref{eq:Lconstraint1}, \eqref{eq:Lconstraint2}, \eqref{eq:Lconstraint3}, \eqref{eq:Lconstraint4} and
	\begin{align}\label{eq:PLconstraint3}
		\mathbf{A}\boldsymbol{\alpha}+\mathbf{B}\boldsymbol{\beta}\leq\mathbf{c}.
	\end{align}
\end{LP}
LP contains a total of $4n+\ell$ scalar equality and inequality constraints (not including nonnegativity constraints on variables) and $n^2+n+2m$ choice variables, including the $2m$ portfolio weights in $\boldsymbol{\alpha}$ and $\boldsymbol{\beta}$ we seek to compute plus an additional $n^2+n$ auxilliary choice variables in $\boldsymbol{\Psi}$ and $\boldsymbol{\xi}$. The auxilliary choice variables do not enter the objective function in LP but influence the choice of $\boldsymbol{\alpha}$ and $\boldsymbol{\beta}$ through the constraints. We could reduce the number of scalar equality and inequality constraints in LP from $4n+\ell$ to $3n+\ell$ by dropping the auxiliary choice variable $\boldsymbol{\xi}$ and replacing \eqref{eq:Lconstraint2} and \eqref{eq:Lconstraint3} with
\begin{align}\label{eq:combinedconstraint}
	\mathbf{S}\boldsymbol{\Psi}^\top\boldsymbol{\mu}&\leq\mathbf{S}\boldsymbol{\mu}.
\end{align}
However, as discussed in \citet{Luedtke08}, it is computationally advantageous to use the formulation in LP because \eqref{eq:Lconstraint2} and \eqref{eq:Lconstraint3} together contain at most $n^2+n(n-1)/2+n$ nonzero coefficients, compared to at most $n^2(n-1)/2$ nonzero coefficients in \eqref{eq:combinedconstraint}, an order of magnitude greater.

The solution for $\boldsymbol{\alpha}$ and $\boldsymbol{\beta}$ obtained using LP is identical to the solution obtained using the linear program proposed in \citet{PostLongarela21}. However, the formulation of the latter program is different to LP, and involves a total of $n^2+n+\ell$ scalar equality and inequality constraints, compared to only $4n+\ell$ in LP. We discuss the difference between the two linear programs more fully in Appendix \ref{sec:LPstar}. Results reported there indicate that the runtime required to solve the linear program in \citet{PostLongarela21} can be one or more orders of magnitude greater than the runtime required to solve LP.

\subsection{First-order stochastic dominance}

We now turn to solving the portfolio choice problem posed in Section \ref{sec:problem} when the symbol $\gtrsim$ appearing in the stochastic dominance constraint \eqref{eq:sdconstraint} is understood to denote first-order stochastic dominance. While second-order stochastic dominance constraints may be formulated in terms of linear inequalities and are therefore convenient to implement in a linear program, first-order stochastic dominance constraints do not generally admit such a formulation. They may instead be formulated in terms of a combination of linear inequalities and binary constraints. Such a formulation was introduced in \citet{Kuosmanen04} and subsequently improved in \citet{Luedtke08}. Theorem 4.1 in the latter article, and the subsequent discussion, implies that any given $(\boldsymbol{\alpha},\boldsymbol{\beta})\in\mathcal{P}$ satisfies the first-order stochastic dominance constraint \eqref{eq:sdconstraint} if and only if there exists an $n\times n$ binary matrix $\boldsymbol{\Psi}$ and an $n\times1$ nonnegative vector $\boldsymbol{\xi}$ satisfying \eqref{eq:Lconstraint1}, \eqref{eq:Lconstraint2}, \eqref{eq:Lconstraint3} and \eqref{eq:Lconstraint4}. This leads us to propose the following mixed-integer linear program for solving our portfolio choice problem.
\begin{MILP}
	Choose two $m\times1$ nonnegative vectors $\boldsymbol{\alpha},\boldsymbol{\beta}$, an $n\times1$ nonnegative vector $\boldsymbol{\xi}$, and an $n\times n$ binary matrix $\boldsymbol{\Psi}$ to maximize $-\mathbf{p}^\top\boldsymbol{\alpha}+\mathbf{q}^\top\boldsymbol{\beta}$ subject to the constraints \eqref{eq:Lconstraint1}, \eqref{eq:Lconstraint2}, \eqref{eq:Lconstraint3}, \eqref{eq:Lconstraint4} and \eqref{eq:PLconstraint3}.
\end{MILP}

The only difference between LP and MILP is that the latter requires the entries of $\boldsymbol{\Psi}$ to be binary. For this reason LP is called the linear relaxation of MILP. The presence of binary constraints on the $n^2$ auxiliary choice variables in $\boldsymbol{\Psi}$ makes MILP much more difficult to solve than LP. General purpose optimizers for mixed-integer linear programs---leading implementations include Gurobi and CPLEX---can be used if one is willing to accept the best feasible solution found within a prespecified maximum runtime.

A judicious choice of starting values can greatly reduce the runtime needed to produce satisfactory solutions to MILP. In our empirical implementation we solve MILP by supplying Gurobi with feasible starting values chosen using a procedure based on Algorithm 1 in \citet{Luedtke08}. We provide a complete description of this procedure, which we call MILP-ST, in Appendix \ref{sec:MILP}. Starting value selection using MILP-ST involves solving a series of relatively small linear programs, and takes less than one second for each of the 215 dates at which we select portfolios. We have found that allowing Gurobi nine seconds of runtime to solve MILP with the MILP-ST starting values delivers results comparable to those obtained by allowing Gurobi five minutes of runtime to solve MILP with the default starting values. In fact, the MILP-ST starting values are typically not much worse (and sometimes better) than the best feasible solution identified by Gurobi after five minutes of runtime with the default starting values, despite requiring less than a second to compute. See Appendix \ref{sec:MILP} for further details. In the following section we solve MILP by supplying Gurobi with the MILP-ST starting values and accepting the best feasible solution found after a maximum runtime of nine seconds.

\section{Empirical implementation}\label{sec:empirical}

\subsection{Data}\label{sec:data}

\begin{table*}[b!]
	\centering
	\begin{threeparttable}
		\caption{Median strike quantities and quote sizes}\label{table:datasummary}
		\begin{tabular}{cccccccccccccc}
			\toprule
			&\multicolumn{6}{c}{Moderate strike range}&&\multicolumn{6}{c}{Wide strike range}\\\cmidrule{2-7}\cmidrule{9-14}
			&&\multicolumn{5}{c}{Median quote size}&&&\multicolumn{5}{c}{Median quote size}\\\cmidrule{3-7}\cmidrule{10-14}
			&&\multicolumn{2}{c}{Call}&&\multicolumn{2}{c}{Put}&&&\multicolumn{2}{c}{Call}&&\multicolumn{2}{c}{Put}\\\cmidrule{3-4}\cmidrule{6-7}\cmidrule{10-11}\cmidrule{13-14}
			Years&Med.\ strikes&Bid&Ask&&Bid&Ask&&Med.\ strikes&Bid&Ask&&Bid&Ask\\
			\midrule
			2004--06&33&50&50&&50&50&&55&50&50&&50&50\\
			2007--09&38&54&54&&53&53&&80&54&54&&54&54\\
			2010--12&38&200&172&&185&177&&109&140&140&&113&112\\
			2013--15&59&114&112&&103&103&&151&114&112&&100&100\\
			2016--18&73&51&54&&190&186&&182&50&50&&174&190\\
			2019--21&97&120&100&&222&195&&258&115&115&&468&400\\
			\bottomrule
		\end{tabular}
		
	\end{threeparttable}
	\floatfoot{\emph{Notes:} The table shows the median number of strikes at which options are written, and the median quote size at each best bid and ask price, for each 36 month period from 2004 to 2021 (35 months in 2019--21). The moderate strike range includes strikes between 10\% below and 5\% above the current SPX, and the wide strike range includes strikes between 30\% below and 15\% above.}
\end{table*}

The SPX options data for our empirical implementation were purchased from the CBOE DataShop and span the 215 months from January 2004 to November 2021. In each of those months prior to January 2015 (following December 2014) we confine our attention to put and call options with the standard classification codes SPX, SPT, SPQ, SPZ, SXB, SXM, SXY, SXZ, SZP and with 29 (28) days to maturity. These options are all traded on the same day in a given month and expire on the third Saturday (Friday) of the following month. In 8 of the 215 months the choice of dates differs slightly from the description just given, generally due to a public holiday, but the time to expiry is always 28, 29 or 30 days. For each option we observe the best bid and ask prices quoted at 2:45pm Central Time (30 minutes before closing) and the associated bid and ask sizes (i.e., volumes). Table \ref{table:datasummary} provides summary statistics for the numbers of strikes and the quote sizes for 36 month subperiods.

At each date on which portfolios are chosen we use the LIBOR as the risk-free rate and use dividend yield data obtained from the multpl online database (\href{www.multpl.com}{www.multpl.com}) to impute a forward price for delivery of one unit of the SPX at the time of expiry. Excess returns for the market portfolio are calculated under the assumption that this portfolio is purchased at the discounted forward rate and liquidated at the expiry date. Excess returns for portfolios including options are calculated under the assumption that option premia are invested at the risk-free rate and liquidated at the expiry date. Our method of imputing forward prices does not affect the performance of the market portfolio relative to the performance of the market portfolio enhanced with the selected layover portfolio, because both include a single unit of the SPX.

Daily realized variances are used to determine the variance risk premium used to form our specification of the SPX return distribution. We computed daily realized variances using historical 5 minute SPX data obtained from LSEG Data \& Analytics. Following \citet[p.~10]{BollerslevTodorov23} we computed the realized variance for each day by summing the squared five minute returns observed during the trading day to obtain an intraday realized variance, and then multiplying this intraday variance by an overnight adjustment factor based on the average ratio of the squared overnight return and intraday realized variance over the past year.

We eliminate pure arbitrage opportunities from our option price data by removing contracts which imply a negatively priced vertical spread or butterfly spread. Such contracts are identified by checking the inequalities provided in \cite{CarrMadan05}. A total of 37 contracts are removed, all in the years 2004--2006. Of these 37 contracts, 26 are written at strikes falling outside our moderate strike range spanning from 10\% below the current SPX to 5\% above. While pure arbitrage opportunities satisfy our definition of stochastic arbitrage, we exclude them to focus attention on those stochastic arbitrage opportunities not providing a riskless profit. For part of our analysis in Section \ref{sec:resultswide} we restore the pure arbitrage opportunities to our data set.

As in \citet{PostLongarela21}, we require short and long positions in options to be taken at the respective bid and ask prices but do not otherwise account for transaction costs in our analysis. There are two justifications for this. First, the inclusion of transaction costs would only strengthen our finding that LP and MILP do not in fact identify stochastic arbitrage opportunities with profitable realized performance. Second, as documented by \citet{MuravyevPearson20}, the bid-ask spread may actually overstate the effective cost of trading options for algorithmic traders who can time transactions so as to trade within the spread.

\subsection{Market depth constraint}\label{sec:marketdepth}

The quoted bid and ask sizes for each option are used to determine $\mathbf{v}$ and $\mathbf{w}$, the vectors of maximum long and short positions, which may be described as providing market depth constraints. How this is done depends on the size of the assumed unit investment in the SPX. Following \citet{PostLongarela21} we consider four sizes of the unit investment: $S=1,10,100,1000$. The entries of $\mathbf{v}$ and $\mathbf{w}$ are set equal to $1/S$ times the quoted ask and bid sizes respectively. Since the multiplier of SPX options is \$100 per index point, defining $\mathbf{v}$ and $\mathbf{w}$ in this way means that a unit investment in the SPX corresponds to an investment of $\$100S$ times the SPX. The average of the SPX over the 215 dates at which option portfolios are chosen is 1900. We may therefore consider the four sizes $S=1,10,100,1000$ as corresponding roughly to market investments of \$190,000, \$1.9 million, \$19 million and \$190 million respectively. As discussed in \citet{PostLongarela21}, the market depth constraints are on the one hand permissive because the investor is assumed to be able to buy or sell all options listed at the best ask or bid price before other traders may intervene, but are on the other hand restrictive because the investor is unable to buy or sell options listed above the best ask price or below the best bid price.

\subsection{Specification of state probabilities}\label{sec:phys}

The programs LP and MILP require us to specify state probabilities for the SPX value at the time of option expiry. In each month we specify location and scale parameters for the SPX return at expiry. The location parameter is equal to the current risk-free rate plus a constant market risk premium, while the scale parameter is equal to the current one-month-ahead option implied volatility (as determined by the VIX) divided by a constant factor. The SPX return distribution, normalized by location and scale, is assumed to be a member of the family of skewed generalized t-distributions (SGT distributions) introduced in \cite{Theodossiou98}. We discretize the continuous SPX return distribution to produce state probabilities for LP and MILP in increments of 5 index points over the range of strikes.

A range of values for the market risk premium have been used in prior empirical studies. \citet[Online Appendix 17.5, p.\ 10]{LutolfCarrollPirnes09} argue that an annualized market risk premium of between 5\% and 7\% is reasonable for US equities. We set the market risk premium equal to 6.67\%, which is the annualized mean one-month SPX excess return observed at daily frequency over our sample period of 2004--2021. Reducing the market risk premium to 5\% or increasing it to 8\% does not qualitatively affect our empirical findings. \citet[p.~104]{PostLongarela21} remark that the choice of fixed market risk premium is largely inconsequential for portfolio selection with LP.

As is well known, option implied volatility measures such as the VIX tend to provide an overestimate of realized volatility and should therefore be deflated in some way to produce a more accurate forward-looking measure of realized volatility. While this is commonly done by subtracting a constant variance risk premium from the square of option implied volatility, \citet[p.~309]{ProkopczukSimen14} argue that a multiplicative variance risk premium is generally more stable over time and ought to be preferred to an additive correction. Following this advice, we use a constant multiplicative variance risk premium for our analysis. We calculate it by averaging the daily ratio of the one-month-ahead option implied variance (as determined by the VIX) and corresponding realized variance (obtained by summing daily realized variances) over our sample period of 2004-2021. This produces a multiplicative variance risk premium of $1.41^2$. The scale parameter for our specification of the SPX return distribution is obtained by dividing the current one-month-ahead option implied volatility by 1.41.

Our specification of state probabilities assumes that the SPX monthly return distribution, normalized by location and scale as just described, is a member of the SGT family. As discussed in \cite{Theodossiou98}, members of this family are determined by three parameters: shape ($k$), degrees of freedom ($\nu$) and asymmetry ($\lambda$). The standard normal distribution is obtained by setting $k=2$, $\nu=\infty$ and $\lambda=0$. We computed maximum likelihood estimates of $k$, $\nu$ and $\lambda$ using daily observations of normalized monthly SPX returns over our sample period, constraining the estimate of $\nu$ to be no less than 5, the minimum integer value required to obtain a finite fourth moment. The computed maximum likelihood estimates were $\hat{k}=1.85$, $\hat{\nu}=5$ and $\hat{\lambda}=-0.53$. Panel (a) of Figure \ref{fig:PDFs} displays the estimated SGT distribution overlaid on a histogram of normalized monthly SPX returns. The standard normal distribution is also displayed. Plainly, the estimated SGT distribution provides a much better fit to the data than does the standard normal distribution, with the latter failing to capture the asymmetry of the SPX return distribution.
\begin{figure*}
	\centering
	\subcaptionbox{Normalized SPX return histogram}{
		\centering
		\begin{tikzpicture}[scale=3.4]
			\draw[black, thick,->] (0,0) -- (0,1.1);
			\draw[black, thick] (0,0) -- (1.5,0);
			\draw (0,0.3pt) -- (0,-1pt)%
			node[anchor=north,font=\scriptsize] {$-3$};
			\draw (0.3,0.3pt) -- (0.3,-1pt)%
			node[anchor=north,font=\scriptsize] {$-2$};
			\draw (0.6,0.3pt) -- (0.6,-1pt)%
			node[anchor=north,font=\scriptsize] {$-1$};
			\draw (0.9,0.3pt) -- (0.9,-1pt)%
			node[anchor=north,font=\scriptsize] {$0$};
			\draw (1.2,0.3pt) -- (1.2,-1pt)%
			node[anchor=north,font=\scriptsize] {$1$};
			\draw (1.5,0.3pt) -- (1.5,-1pt)%
			node[anchor=north,font=\scriptsize] {$2$};
			\draw (0.3pt,0) -- (-0.8pt,0);
			\draw (0.3pt,{1/4}) -- (-0.8pt,{1/4});
			\draw (0.3pt,{2/4}) -- (-0.8pt,{2/4});
			\draw (0.3pt,{3/4}) -- (-0.8pt,{3/4});
			\draw (0.3pt,{4/4}) -- (-0.8pt,{4/4});
			\node[left,font=\scriptsize] at (0,0) {$0$};
			\node[left,font=\scriptsize] at (0,{1/4}) {$.05$};
			\node[left,font=\scriptsize] at (0,{2/4}) {$.1$};
			\node[left,font=\scriptsize] at (0,{3/4}) {$.15$};
			\node[left,font=\scriptsize] at (0,{4/4}) {$.2$};
			\draw[fill=blue!15] (0,0) rectangle (.1,{0.0067/.2});
			\draw[fill=blue!15] (.1,0) rectangle (.2,{0.0113/.2});
			\draw[fill=blue!15] (.2,0) rectangle (.3,{0.0173/.2});
			\draw[fill=blue!15] (.3,0) rectangle (.4,{0.0242/.2});
			\draw[fill=blue!15] (.4,0) rectangle (.5,{0.0327/.2});
			\draw[fill=blue!15] (.5,0) rectangle (.6,{0.0540/.2});
			\draw[fill=blue!15] (.6,0) rectangle (.7,{0.0614/.2});
			\draw[fill=blue!15] (.7,0) rectangle (.8,{0.0778/.2});
			\draw[fill=blue!15] (.8,0) rectangle (.9,{0.1136/.2});
			\draw[fill=blue!15] (.9,0) rectangle (1,{0.1470/.2});
			\draw[fill=blue!15] (1,0) rectangle (1.1,{0.1681/.2});
			\draw[fill=blue!15] (1.1,0) rectangle (1.2,{0.1476/.2});
			\draw[fill=blue!15] (1.2,0) rectangle (1.3,{0.0845/.2});
			\draw[fill=blue!15] (1.3,0) rectangle (1.4,{0.0278/.2});
			\draw[fill=blue!15] (1.4,0) rectangle (1.5,{0.0080/.2});
			\draw[red,thick] plot file {sgtpdf.txt};
			\draw[blue,thick,dashed] plot file {normpdf.txt};
			\draw[red,thick] plot file {sgtpdf.txt};			
		\end{tikzpicture}
		\hspace{.4cm}
	}
	\subcaptionbox{PIT histogram: SGT}{
		\centering
		\begin{tikzpicture}[scale=3.4]
			\draw[black, thick,->] (0,0) -- (0,1.1);
			\draw[black, thick] (0,0) -- (1,0);
			\draw (0,0.3pt) -- (0,-1pt)%
			node[anchor=north,font=\scriptsize] {$0$};
			\draw (.2,0.3pt) -- (.2,-1pt)%
			node[anchor=north,font=\scriptsize] {$.2$};
			\draw (.4,0.3pt) -- (.4,-1pt)%
			node[anchor=north,font=\scriptsize] {$.4$};
			\draw (.6,0.3pt) -- (.6,-1pt)%
			node[anchor=north,font=\scriptsize] {$.6$};
			\draw (.8,0.3pt) -- (.8,-1pt)%
			node[anchor=north,font=\scriptsize] {$.8$};
			\draw (1,0.3pt) -- (1,-1pt)%
			node[anchor=north,font=\scriptsize] {$1$};
			\draw (0.3pt,0) -- (-0.8pt,0);
			\draw (0.3pt,{1/4}) -- (-0.8pt,{1/4});
			\draw (0.3pt,{2/4}) -- (-0.8pt,{2/4});
			\draw (0.3pt,{3/4}) -- (-0.8pt,{3/4});
			\draw (0.3pt,{4/4}) -- (-0.8pt,{4/4});
			\node[left,font=\scriptsize] at (0,0) {$0$};
			\node[left,font=\scriptsize] at (0,{1/4}) {$.05$};
			\node[left,font=\scriptsize] at (0,{2/4}) {$.1$};
			\node[left,font=\scriptsize] at (0,{3/4}) {$.15$};
			\node[left,font=\scriptsize] at (0,{4/4}) {$.2$};	
			\draw[fill=blue!15] (0,0) rectangle (.1,{.1265/.2});	
			\draw[fill=blue!15] (.1,0) rectangle (.2,{.1029/.2});	
			\draw[fill=blue!15] (.2,0) rectangle (.3,{.0860/.2});	
			\draw[fill=blue!15] (.3,0) rectangle (.4,{.0912/.2});	
			\draw[fill=blue!15] (.4,0) rectangle (.5,{.0907/.2});	
			\draw[fill=blue!15] (.5,0) rectangle (.6,{.0912/.2});	
			\draw[fill=blue!15] (.6,0) rectangle (.7,{.0920/.2});	
			\draw[fill=blue!15] (.7,0) rectangle (.8,{.0923/.2});	
			\draw[fill=blue!15] (.8,0) rectangle (.9,{.0912/.2});	
			\draw[fill=blue!15] (.9,0) rectangle (1,{.1361/.2});
			\draw[dashed] (0,.5) -- (1,.5);
		\end{tikzpicture}
		\hspace{.4cm}
	}
	\subcaptionbox{PIT histogram: $N(0,1)$}{
		\centering
		\begin{tikzpicture}[scale=3.4]
			\draw[black, thick,->] (0,0) -- (0,1.1);
			\draw[black, thick] (0,0) -- (1,0);
			\draw (0,0.3pt) -- (0,-1pt)%
			node[anchor=north,font=\scriptsize] {$0$};
			\draw (.2,0.3pt) -- (.2,-1pt)%
			node[anchor=north,font=\scriptsize] {$.2$};
			\draw (.4,0.3pt) -- (.4,-1pt)%
			node[anchor=north,font=\scriptsize] {$.4$};
			\draw (.6,0.3pt) -- (.6,-1pt)%
			node[anchor=north,font=\scriptsize] {$.6$};
			\draw (.8,0.3pt) -- (.8,-1pt)%
			node[anchor=north,font=\scriptsize] {$.8$};
			\draw (1,0.3pt) -- (1,-1pt)%
			node[anchor=north,font=\scriptsize] {$1$};
			\draw (0.3pt,0) -- (-0.8pt,0);
			\draw (0.3pt,{1/4}) -- (-0.8pt,{1/4});
			\draw (0.3pt,{2/4}) -- (-0.8pt,{2/4});
			\draw (0.3pt,{3/4}) -- (-0.8pt,{3/4});
			\draw (0.3pt,{4/4}) -- (-0.8pt,{4/4});
			\node[left,font=\scriptsize] at (0,0) {$0$};
			\node[left,font=\scriptsize] at (0,{1/4}) {$.05$};
			\node[left,font=\scriptsize] at (0,{2/4}) {$.1$};
			\node[left,font=\scriptsize] at (0,{3/4}) {$.15$};
			\node[left,font=\scriptsize] at (0,{4/4}) {$.2$};	
			\draw[fill=blue!15] (0,0) rectangle (.1,{.1149/.2});	
			\draw[fill=blue!15] (.1,0) rectangle (.2,{.0747/.2});	
			\draw[fill=blue!15] (.2,0) rectangle (.3,{.0643/.2});	
			\draw[fill=blue!15] (.3,0) rectangle (.4,{.0656/.2});	
			\draw[fill=blue!15] (.4,0) rectangle (.5,{.0943/.2});	
			\draw[fill=blue!15] (.5,0) rectangle (.6,{.1087/.2});	
			\draw[fill=blue!15] (.6,0) rectangle (.7,{.1345/.2});	
			\draw[fill=blue!15] (.7,0) rectangle (.8,{.1592/.2});	
			\draw[fill=blue!15] (.8,0) rectangle (.9,{.1372/.2});	
			\draw[fill=blue!15] (.9,0) rectangle (1,{.0467/.2});
			\draw[dashed] (0,.5) -- (1,.5);
		\end{tikzpicture}
		\hspace{.4cm}
	}
	\caption{Goodness-of-fit of specified SPX return distributions}\label{fig:PDFs}
	\floatfoot{\emph{Notes:} Panel (a) displays a histogram of daily observations of location- and scale-normalized one-month SPX returns during 2004--2021, overlaid with density functions for the $N(0,1)$ and skewed generalized t (SGT) distributions, with parameters for the latter estimated by maximum likelihood. Panels (b) and (c) display histograms for the probability integral transforms (PITs) associated with the two density functions.}
\end{figure*}

To further investigate the fit of our specification of state probabilities we applied the approach discussed in \citet{DieboldGuntherTay98}, which involves assessing the uniformity of probability integral transforms (PITs). For each date $t$ in our sample period we computed the PIT $F_t(x_{t})$, where $x_t$ is the value of the SPX at date $t$, and $F_t$ is the cumulative distribution function used to predict $x_t$ one month prior. Thus $F_t$ depends on the risk-free rate and VIX one month prior to date $t$, the assumed market and variance risk premia, and the estimated parameters of the SGT distribution. Panel (b) in Figure \ref{fig:PDFs} displays the proportion of PITs belonging to each decile of the unit interval. If state probabilities are correctly specified then the PITs should be distributed uniformly over the unit interval, so that the proportion of PITs belonging to each decile is close to 0.1. We see that the proportions are reasonably close to 0.1 at all deciles except the top and bottom deciles, where the proportions are closer to 0.13. This discrepancy indicates that very large positive or negative SPX returns occur more frequently than is predicted by our specified return distribution. The problem is of lesser importance for our results pertaining to options written at moderate strikes, because the specification of state probabilities outside the permitted range of strikes is irrelevant to our choice of option portfolios. Only the relative probabilities assigned to states within the range of strikes are relevant. We will see, however, that the underprediction of the incidence of very large SPX returns significantly affects the results we obtain with the wide range of strikes, reported in Section \ref{sec:resultswide}.

Panel (c) in Figure \ref{fig:PDFs} shows how the PIT histogram displayed in panel (b) is affected if one uses the standard normal distribution in place of the estimated SGT distribution. It is apparent that using the standard normal distribution leads to a specification of state probabilities which substantially overpredicts the incidence of moderate negative returns, and substantially underpredicts the incidence of moderate positive returns. We will come back to this point toward the end of the following section.

We have explored other specifications of state probabilities which permit more flexible variation over time. Some of these are discussed in Appendix \ref{sec:tv}. Replacing the constant multiplicative variance risk premium with a time-varying multiplicative variance risk premium computed as in \citet{ProkopczukSimen14} did not greatly affect the results of our analysis. Neither did permitting time-variation in the SGT skewness parameter using the score-driven filter introduced in \citet{LangeOsDijk24}. We also tried introducing time-variation in the market risk premium by allowing it vary linearly with an estimate of the additive variance risk premium, as in \citet{BollerslevTauchenZhou09} and \citet{BekaertHoerova14}. This produced a very noisy estimate of the market risk premium, negatively affecting the performance of selected portfolios.

\subsection{Results: moderate strike range}\label{sec:results}

We first report the outcome of our search for stochastic arbitrage opportunities among options written at the moderate range of strikes between 10\% below and 5\% above the current SPX value. Table \ref{table:mainresults} reports selected characteristics of the option portfolios selected using LP and MILP, including realized performance measures. The top half of the table reports the results obtained using the SGT specification of state probabilities described in Section \ref{sec:phys}, while the bottom half reports the results obtained using the standard normal distribution in place of the fitted SGT distribution. The latter results are reported to illustrate what can go wrong when a poor specification of state probabilities is used, and to draw a connection to past literature on the pricing kernel puzzle.

The most important finding in the top half of Table \ref{table:mainresults} is contained in the two lines titled ``Pctg.\ premia $>0.1\%$ mkt.\ investment''. These lines report the percentage of the 215 option portfolio formation dates in which the premium generated by the selected layover portfolio exceeds $0.1\%$ of the value of the unit investment in the SPX. We choose $0.1\%$ as a threshold because if premia are consistently below this level then the option trading strategy is not expected to add much more than a single annualized percentage point to the return of the unit investment in the SPX. Our headline finding is that, using the SGT specification, premia rarely exceed this threshold. The percentage of months in which the threshold is exceeded is 5.1\% (i.e., 11 months) with LP and the unit scale constraint ($S=1$), falling to 2.8\% with the tightest scale constraint ($S=1000$). It is natural to see a decline as we increase $S$ because tightening the market depth constraint reduces the value of tradable options as a proportion of the market investment. With MILP the threshold is exceeded even more rarely, and not at all with the tightest scale constraint. The realized mean excess return generated by the enhanced portfolios (including the premium generated) is, for both LP and MILP and for all scale constraints, slightly lower than the realized mean SPX excess return of 9.81\% over the 215 months in our sample. Table \ref{table:mainresults} reports p-values for simple t-tests of the null hypothesis that the SPX and enhanced portfolio have equal mean excess return, and for tests of the null hypothesis that the enhanced portfolio excess return stochastically dominates (in the second-order sense for LP, and in the first-order sense for MILP) the SPX excess return. The tests of stochastic dominance are based on uniformly weighted Cram\'{e}r-von Mises statistics, as in \cite{LintonSongWhang10}. The p-values do not indicate significant violations of these hypotheses, reflecting the fact that LP and MILP rarely select portfolios which differ substantially from the market portfolio.

\begin{table*}
	\centering
	\begin{threeparttable}
		\caption{Characteristics of selected portfolios with moderate strike range}\label{table:mainresults}
		\begin{tabular}{rrllccccc}
			\toprule
			\multicolumn{5}{c}{}&\multicolumn{4}{c}{Enhanced portfolio}\\\cmidrule{6-9}
			\multicolumn{4}{c}{}& SPX & $S=1$ & $S=10$ & $S=100$ & $S=1000$\\
			\midrule
			\multirow{22}{*}{\rotatebox[origin=c]{90}{Normalized return distribution: skewed generalized t}}&\multirow{11}{*}{\rotatebox[origin=c]{90}{LP}}&Pctg.\ premia $>0.1\% $ mkt.\ investment&&&  5.1 &  4.7 &  3.7 &  2.8 \\
			&&Value of options as a percentage of&Calls bought&&  0.8 &  0.4 &  0.3 &  0.2 \\
			&&$\quad$market investment (average)&Calls written&&  0.8 &  0.5 &  0.5 &  0.3 \\
			&&&Puts bought&&  0.5 &  0.4 &  0.4 &  0.3 \\
			&&&Puts written&&  0.6 &  0.3 &  0.3 &  0.2 \\
			&&Realized moments of excess returns&Mean&9.81& 9.36 & 9.18 & 9.15 & 9.26 \\
			&&&Std.\ dev.&17.77& 17.71 & 17.72 & 17.73 & 17.73 \\
			&&&Skew&-1.58& -1.57 & -1.56 & -1.55 & -1.56 \\
			&&&Sortino&0.71& 0.68 & 0.67 & 0.67 & 0.67 \\
			&&Stochastic dominance p-values&&& 0.320 & 0.219 & 0.215 & 0.218 \\
			&&Equal mean excess return p-value&&& 0.503 & 0.335 & 0.308 & 0.271 \\
			
			\cmidrule{2-9}
			&\multirow{11}{*}{\rotatebox[origin=c]{90}{MILP}}&Pctg.\ premia $>0.1\% $ mkt.\ investment&&&  1.9 &  1.4 &  0.9 &  0.0 \\
			&&Value of options as a percentage of&Calls bought&&  0.6 &  0.2 &  0.1 &  0.0 \\
			&&$\quad$market investment (average)&Calls written&&  0.4 &  0.2 &  0.1 &  0.0 \\
			&&&Puts bought&&  0.3 &  0.1 &  0.1 &  0.0 \\
			&&&Puts written&&  0.5 &  0.1 &  0.1 &  0.0 \\
			&&Realized moments of excess returns&Mean&9.81& 9.63 & 9.52 & 9.51 & 9.81 \\
			&&&Std.\ dev.&17.77& 17.78 & 17.78 & 17.79 & 17.77 \\
			&&&Skew&-1.58& -1.57 & -1.57 & -1.57 & -1.58 \\
			&&&Sortino&0.71& 0.70 & 0.69 & 0.69 & 0.71 \\
			&&Stochastic dominance p-values&&& 0.196 & 0.139 & 0.146 & 0.540 \\
			&&Equal mean excess return p-value&&& 0.426 & 0.178 & 0.196 & 0.318 \\
			
			\midrule
			\multirow{22}{*}{\rotatebox[origin=c]{90}{Normalized return distribution: standard normal}}&\multirow{11}{*}{\rotatebox[origin=c]{90}{LP}}&Pctg.\ premia $>0.1\% $ mkt.\ investment&&& 47.4 & 41.4 & 35.3 & 24.7 \\
			&&Value of options as a percentage of&Calls bought&&  2.0 &  1.3 &  1.0 &  0.6 \\
			&&$\quad$market investment (average)&Calls written&&  3.2 &  2.5 &  2.0 &  1.2 \\
			&&&Puts bought&&  2.3 &  1.9 &  1.6 &  1.1 \\
			&&&Puts written&&  1.8 &  1.2 &  1.1 &  0.8 \\
			&&Realized moments of excess returns&Mean&9.81& 6.65 & 5.53 & 6.58 & 8.24 \\
			&&&Std.\ dev.&17.77& 17.61 & 17.70 & 17.70 & 17.54 \\
			&&&Skew&-1.58& -1.31 & -1.24 & -1.30 & -1.41 \\
			&&&Sortino&0.71& 0.50 & 0.41 & 0.49 & 0.61 \\
			&&Stochastic dominance p-values&&& 0.118 & 0.038 & 0.074 & 0.144 \\
			&&Equal mean excess return p-value&&& 0.205 & 0.064 & 0.103 & 0.204 \\
			
			\cmidrule{2-9}
			&\multirow{11}{*}{\rotatebox[origin=c]{90}{MILP}}&Pctg.\ premia $>0.1\% $ mkt.\ investment&&& 24.2 & 22.3 & 20.5 &  9.3 \\
			&&Value of options as a percentage of&Calls bought&&  1.7 &  1.0 &  0.9 &  0.3 \\
			&&$\quad$market investment (average)&Calls written&&  2.1 &  1.6 &  1.4 &  0.5 \\
			&&&Puts bought&&  1.4 &  1.2 &  1.2 &  0.5 \\
			&&&Puts written&&  1.3 &  0.9 &  0.9 &  0.3 \\
			&&Realized moments of excess returns&Mean&9.81& 7.58 & 7.90 & 7.13 & 9.63 \\
			&&&Std.\ dev.&17.77& 17.82 & 17.87 & 17.79 & 17.68 \\
			&&&Skew&-1.58& -1.36 & -1.36 & -1.35 & -1.50 \\
			&&&Sortino&0.71& 0.55 & 0.58 & 0.52 & 0.71 \\
			&&Stochastic dominance p-values&&& 0.132 & 0.149 & 0.066 & 0.401 \\
			&&Equal mean excess return p-value&&& 0.204 & 0.259 & 0.061 & 0.826 \\
			\bottomrule
		\end{tabular}
	\end{threeparttable}
	\floatfoot{\emph{Notes:} The table reports characteristics of the 215 portfolios selected with LP or with MILP, using the fitted SGT (top half) or standard normal (bottom half) specifications of normalized SPX returns. The range of included strikes is from 10\% below to 5\% above the current SPX. The realized mean and standard deviation of portfolio returns are reported in annualized percentage points. The reported p-values correspond to the null hypothesis that the enhanced portfolio excess returns stochastically dominate the SPX excess returns (in the second order sense for LP, and in the first order sense for MILP) and to the null hypothesis that the enhanced portfolio excess returns and SPX excess returns have equal mean.}
\end{table*}

\begin{figure*}
	\caption{Layover portfolio payoffs with normal specification of SPX return distributions}\label{fig:Normalpayoffs}
	\centering
	\subcaptionbox{Portfolios selected with LP}{
		\centering
		\begin{tikzpicture}[scale=3.8]
			\draw[black, thick,<->] (0,-.6) -- (0,.6);
			\draw[black, thick,<->] (-.1,-.5) -- (1.5,-.5);
			\draw ({(.9-.875)*1.5/(1.075-.875)},{-.5cm+0.3pt}) -- ({(.9-.875)*1.5/(1.075-.875)},{-.5cm-1pt})
			node[anchor=north,font=\scriptsize] {$.9$};
			\draw ({(.95-.875)*1.5/(1.075-.875)},{-.5cm+0.3pt}) -- ({(.95-.875)*1.5/(1.075-.875)},{-.5cm-1pt})
			node[anchor=north,font=\scriptsize] {$.95$};
			\draw ({(1-.875)*1.5/(1.075-.875)},{-.5cm+0.3pt}) -- ({(1-.875)*1.5/(1.075-.875)},{-.5cm-1pt})
			node[anchor=north,font=\scriptsize] {$1$};
			\draw ({(1.05-.875)*1.5/(1.075-.875)},{-.5cm+0.3pt}) -- ({(1.05-.875)*1.5/(1.075-.875)},{-.5cm-1pt})
			node[anchor=north,font=\scriptsize] {$1.05$};
			\draw (0.3pt,{-.1*.6/.14}) -- (-0.8pt,{-.1*.6/.14});
			\draw (0.3pt,{-.05*.6/.14}) -- (-0.8pt,{-.05*.6/.14});
			\draw (0.3pt,0) -- (-0.8pt,0);
			\draw (0.3pt,{.05*.6/.14}) -- (-0.8pt,{.05*.6/.14});
			\draw (0.3pt,{.1*.6/.14}) -- (-0.8pt,{.1*.6/.14});
			\node[left,font=\scriptsize] at (0,{-.1*.6/.14}) {$-.1$};
			\node[left,font=\scriptsize] at (0,{-.05*.6/.14}) {$-.05$};
			\node[left,font=\scriptsize] at (0,0) {$0$};
			\node[left,font=\scriptsize] at (0,{.05*.6/.14}) {$.05$};
			\node[left,font=\scriptsize] at (0,{.1*.6/.14}) {$.1$};
			\draw[thin] (0,0) -- (1.5,0);
			\draw[thick,only marks,olive] plot[mark=*,mark size=.15pt] file {realizedpayoffLP.txt};
			\node[font=\footnotesize] at ({3/4},-.725) {SPX at expiry};
			\node[rotate=90,font=\footnotesize] at (-.275,0) {Layover portfolio payoff};
		\end{tikzpicture}
		\hspace{1cm}
	}
	\subcaptionbox{Portfolios selected with MILP}{
		\centering
		\begin{tikzpicture}[scale=3.8]
			\draw[black, thick,<->] (0,-.6) -- (0,.6);
			\draw[black, thick,<->] (-.1,-.5) -- (1.5,-.5);
			\draw ({(.9-.875)*1.5/(1.075-.875)},{-.5cm+0.3pt}) -- ({(.9-.875)*1.5/(1.075-.875)},{-.5cm-1pt})
			node[anchor=north,font=\scriptsize] {$.9$};
			\draw ({(.95-.875)*1.5/(1.075-.875)},{-.5cm+0.3pt}) -- ({(.95-.875)*1.5/(1.075-.875)},{-.5cm-1pt})
			node[anchor=north,font=\scriptsize] {$.95$};
			\draw ({(1-.875)*1.5/(1.075-.875)},{-.5cm+0.3pt}) -- ({(1-.875)*1.5/(1.075-.875)},{-.5cm-1pt})
			node[anchor=north,font=\scriptsize] {$1$};
			\draw ({(1.05-.875)*1.5/(1.075-.875)},{-.5cm+0.3pt}) -- ({(1.05-.875)*1.5/(1.075-.875)},{-.5cm-1pt})
			node[anchor=north,font=\scriptsize] {$1.05$};
			\draw (0.3pt,{-.1*.6/.14}) -- (-0.8pt,{-.1*.6/.14});
			\draw (0.3pt,{-.05*.6/.14}) -- (-0.8pt,{-.05*.6/.14});
			\draw (0.3pt,0) -- (-0.8pt,0);
			\draw (0.3pt,{.05*.6/.14}) -- (-0.8pt,{.05*.6/.14});
			\draw (0.3pt,{.1*.6/.14}) -- (-0.8pt,{.1*.6/.14});
			\node[left,font=\scriptsize] at (0,{-.1*.6/.14}) {$-.1$};
			\node[left,font=\scriptsize] at (0,{-.05*.6/.14}) {$-.05$};
			\node[left,font=\scriptsize] at (0,0) {$0$};
			\node[left,font=\scriptsize] at (0,{.05*.6/.14}) {$.05$};
			\node[left,font=\scriptsize] at (0,{.1*.6/.14}) {$.1$};
			\draw[thin] (0,0) -- (1.5,0);
			\draw[thick,only marks,olive] plot[mark=*,mark size=.15pt] file {realizedpayoffMILP.txt};
			\node[font=\footnotesize] at ({3/4},-.725) {SPX at expiry};
			\node[rotate=90,font=\footnotesize] at (-.275,0) {Layover portfolio payoff};
		\end{tikzpicture}
		\hspace{1cm}
	}
	\floatfoot{\emph{Notes:} Each panel displays the realized payoffs of the 215 layover portfolios selected using either LP or MILP when $S=1$ and the normalized SPX return distribution is specified to be standard normal. The horizontal axes measure the SPX on the date of expiry as a proportion of the SPX on the portfolio formation date. The range of included strikes is from 10\% below to 5\% above the current SPX.}
\end{figure*}

The results reported in the bottom half of Table \ref{table:mainresults}, which are obtained by specifying state probabilities using the standard normal distribution rather than the fitted SGT distribution, are very different to those reported in the top half. Layover portfolios generating substantial premia are selected in many more months: depending on the scale parameter $S$, premia exceed 0.1\% of the unit market investment in between 24.7\% and 47.4\% of months using LP, and in between 9.3\% and 24.2\% of months using MILP. The average values of puts and calls bought and written reported in the bottom half of Table \ref{table:mainresults} are substantially higher than those reported in the top half. These larger positions in options do not, however, generate realized profits. For $S=1$, $S=10$ and $S=100$, the realized mean excess returns (including premia) for the enhanced portfolios are more than three percentage points below the realized mean SPX excess return using LP, and two or more percentage points below using MILP, though the differences are only borderline statistically significant, or not significant.

We can learn more about the poor performance of the portfolios selected using the standard normal specification by directly examining the payoffs they deliver. Figure \ref{fig:Normalpayoffs} displays plots of the 215 layover portfolio payoffs against the payoff of the market portfolio, with the payoffs of the layover portfolios selected using LP in panel (a) and those selected using MILP in panel (b). Layover portfolio payoffs of zero, which are frequent, occur when the enhanced portfolio and market portfolio deliver the same payoff. We see that negative layover portfolio payoffs are generally associated with moderately positive market returns, while positive layover portfolio payoffs are generally associated with moderately negative market returns. Moderately positive market returns occur more frequently than moderately negative market returns, dragging down the performance of the layover portfolios. This reflects the misspecification of state probabilities identified in panel (c) of Figure \ref{fig:PDFs}: moderately negative market returns occur less frequently than predicted, while moderately positive market returns occur more frequently than predicted.

The pattern of layover portfolio payoffs displayed in Figure \ref{fig:Normalpayoffs} may shed light on the so-called pricing kernel puzzle. Literature on this puzzle originates with the identification in \cite{Jackwerth00} of an apparent anomaly in SPX option price data, wherein an empirical estimate of the pricing kernel (i.e, the ratio of Arrow security prices to state probabilities) is locally increasing near the center of the SPX return distribution, and decreasing elsewhere. Similar findings were reported in \cite{AitSahaliaLo00} and \cite{RosenbergEngle02}. If the pricing kernel is locally increasing near the center of the SPX return distribution then, given a complete market of Arrow securities, stochastic arbitrage may be achieved by augmenting a unit investment in the SPX with a layover portfolio delivering an N-shaped payoff function which is positive when the SPX return is moderately negative, and negative when the SPX return is moderately positive. This was shown in \cite{Beare11}---see Figure 4.3(b) therein---using the pricing kernel estimated in \cite{Jackwerth00}. The roughly N-shaped layover portfolio payoffs displayed in Figure \ref{fig:Normalpayoffs} therefore resemble the stochastic arbitrage opportunity implied by the nonmonotone shape of the pricing kernel reported in \cite{Jackwerth00}.

The poor performance of the layover portfolios selected using the standard normal specification of state probabilities is due to the failure of this specification to capture the asymmetry at the center of the SPX return distribution. Unmodelled asymmetry may also help to explain the pricing kernel puzzle. While the pricing kernel nonmonotonicity identified in \cite{Jackwerth00} is based on an asymmetric nonparametric estimate of the SPX return distribution, the estimated distribution appears to be roughly symmetric near its center; see Figure 2 therein. Similarly, the nonmonotone pricing kernels displayed in Figures 5 and 6 in \cite{RosenbergEngle02} are based on estimates of the SPX return distribution obtained using an asymmetric GARCH model, but Figure 2 therein shows that the estimated return distributions are roughly symmetric. Unmodelled asymmetry provides a simple explanation for why estimated pricing kernels may be locally increasing at moderate return levels and has, to the best of our knowledge, not being considered in prior literature seeking to explain the pricing kernel puzzle; see, for instance, \cite{HensReichlin13}.

\subsection{Results: wide strike range}\label{sec:resultswide}

The analysis reported in the previous section confined attention to options written at strikes between 10\% below and 5\% above the current SPX. In this section we widen the range of strikes to between 30\% below and 15\% above the current SPX. State probabilities are specified using the SGT distribution as described in Section \ref{sec:phys}. Table \ref{table:wideresults} reports characteristics of the selected option portfolios. The results reported in the top half of Table \ref{table:wideresults} are obtained after eliminating pure arbitrage opportunities from our option price data, while the results reported in the bottom half are based on the full set of option price data including pure arbitrage opportunities.

\begin{table*}
	\centering
	\begin{threeparttable}
		\caption{Characteristics of selected portfolios with wide strike range}\label{table:wideresults}
		
		\begin{tabular}{rrllccccc}
			\toprule
			\multicolumn{5}{c}{}&\multicolumn{4}{c}{Enhanced portfolio}\\\cmidrule{6-9}
			\multicolumn{4}{c}{}& SPX & $S=1$ & $S=10$ & $S=100$ & $S=1000$\\
			\midrule
			\multirow{22}{*}{\rotatebox[origin=c]{90}{Pure arbitrage excluded}}&\multirow{11}{*}{\rotatebox[origin=c]{90}{LP}}&Pctg.\ premia $>0.1\% $ mkt.\ investment&&& 59.1 & 49.8 & 32.1 & 12.1 \\
			&&Value of options as a percentage of&Calls bought&&  1.9 &  1.2 &  1.0 &  0.5 \\
			&&$\quad$market investment (average)&Calls written&&  3.1 &  1.9 &  1.3 &  0.7 \\
			&&&Puts bought&&  3.9 &  0.8 &  1.0 &  0.5 \\
			&&&Puts written&&  3.6 &  0.7 &  1.1 &  0.5 \\
			&&Realized moments of excess returns&Mean&9.81& 7.84 & 6.33 & 6.47 & 7.68 \\
			&&&Std.\ dev.&17.77& 18.08 & 17.80 & 17.79 & 17.42 \\
			&&&Skew&-1.58& -1.58 & -1.59 & -1.61 & -1.61 \\
			&&&Sortino&0.71& 0.55 & 0.45 & 0.46 & 0.57 \\
			&&Stochastic dominance p-values&&& 0.284 & 0.123 & 0.098 & 0.101 \\
			&&Equal mean excess return p-value&&& 0.541 & 0.221 & 0.169 & 0.154 \\
			
			\cmidrule{2-9}
			&\multirow{11}{*}{\rotatebox[origin=c]{90}{MILP}}&Pctg.\ premia $>0.1\% $ mkt.\ investment&&&  2.8 &  1.9 &  1.4 &  0.0 \\
			&&Value of options as a percentage of&Calls bought&&  0.9 &  0.3 &  0.1 &  0.0 \\
			&&$\quad$market investment (average)&Calls written&&  1.4 &  0.4 &  0.1 &  0.0 \\
			&&&Puts bought&&  3.6 &  0.4 &  0.1 &  0.0 \\
			&&&Puts written&&  3.4 &  0.3 &  0.1 &  0.0 \\
			&&Realized moments of excess returns&Mean&9.81& 10.86 & 10.05 & 9.73 & 9.81 \\
			&&&Std.\ dev.&17.77& 17.80 & 17.76 & 17.77 & 17.77 \\
			&&&Skew&-1.58& -1.60 & -1.59 & -1.58 & -1.58 \\
			&&&Sortino&0.71& 0.79 & 0.73 & 0.71 & 0.71 \\
			&&Stochastic dominance p-values&&& 0.945 & 0.754 & 0.153 & 0.971 \\
			&&Equal mean excess return p-value&&& 0.193 & 0.365 & 0.168 & 0.252 \\
			
			\midrule
			\multirow{22}{*}{\rotatebox[origin=c]{90}{Pure arbitrage included}}&\multirow{11}{*}{\rotatebox[origin=c]{90}{LP}}&Pctg.\ premia $>0.1\% $ mkt.\ investment&&& 59.1 & 50.2 & 32.6 & 12.6 \\
			&&Value of options as a percentage of&Calls bought&& 20.3 &  3.0 &  1.2 &  0.5 \\
			&&$\quad$market investment (average)&Calls written&& 21.6 &  3.6 &  1.4 &  0.7 \\
			&&&Puts bought&&  5.0 &  1.0 &  1.0 &  0.5 \\
			&&&Puts written&&  6.2 &  1.1 &  1.1 &  0.5 \\
			&&Realized moments of excess returns&Mean&9.81& 20.93 & 7.73 & 6.62 & 7.70 \\
			&&&Std.\ dev.&17.77& 49.93 & 18.61 & 17.80 & 17.42 \\
			&&&Skew&-1.58& 11.66 & -1.06 & -1.61 & -1.61 \\
			&&&Sortino&0.71& 1.40 & 0.55 & 0.47 & 0.57 \\
			&&Stochastic dominance p-values&&& 0.611 & 0.207 & 0.106 & 0.102 \\
			&&Equal mean excess return p-value&&& 0.408 & 0.511 & 0.190 & 0.157 \\
			
			\cmidrule{2-9}
			&\multirow{11}{*}{\rotatebox[origin=c]{90}{MILP}}&Pctg.\ premia $>0.1\% $ mkt.\ investment&&&  3.3 &  2.3 &  1.9 &  0.5 \\
			&&Value of options as a percentage of&Calls bought&& 19.4 &  2.2 &  0.3 &  0.0 \\
			&&$\quad$market investment (average)&Calls written&& 20.0 &  2.2 &  0.3 &  0.0 \\
			&&&Puts bought&&  4.7 &  0.4 &  0.1 &  0.0 \\
			&&&Puts written&&  6.0 &  0.5 &  0.1 &  0.0 \\
			&&Realized moments of excess returns&Mean&9.81& 24.27 & 11.28 & 9.86 & 9.83 \\
			&&&Std.\ dev.&17.77& 49.63 & 18.44 & 17.79 & 17.77 \\
			&&&Skew&-1.58& 11.71 & -1.18 & -1.58 & -1.58 \\
			&&&Sortino&0.71& 1.66 & 0.81 & 0.71 & 0.71 \\
			&&Stochastic dominance p-values&&& 0.974 & 0.873 & 0.498 & 0.986 \\
			&&Equal mean excess return p-value&&& 0.271 & 0.245 & 0.755 & 0.286 \\
			\bottomrule
		\end{tabular}
		
	\end{threeparttable}
	\floatfoot{\emph{Notes:} The table reports characteristics of the 215 portfolios selected with LP or with MILP, using the SGT distribution to specify state probabilities as discussed in Section \ref{sec:phys}. The results in the top half of the table were computed after eliminating 37 contracts which provided pure arbitrage opportunities, while the results in the bottom half do not exclude these contracts. The range of included strikes is from 30\% below to 15\% above the current SPX. Other details of the table are the same as in Table \ref{table:mainresults}.}
\end{table*}

The results in the top half of Table \ref{table:wideresults} may be compared directly to those in the top half of Table \ref{table:mainresults}, with differences solely attributable to the range of permitted strikes. For MILP, there is not a lot of difference. Substantial stochastic arbitrage opportunities are detected in only a tiny minority of months (though more often than with the moderate strike range), and overall portfolio performance is consequently very similar to that of the market portfolio. It should be noted that the computation of solutions to MILP is difficult using the wide strike range, and there is no guarantee that our computed solutions are optimal. Unlike LP, which is a simple linear program and can therefore be reliably solved, our procedure for solving MILP is to take the best feasible solution identified by Gurobi after a maximum runtime of nine seconds. Tripling the width of the range of strikes increases the number of linear equalities and inequalities in MILP by a factor of roughly three, and increases the numbers of choice variables and binary constraints by a factor of roughly nine. This increased computational complexity may be partly responsible for the sharp discrepancy between the results for LP and for MILP in Table \ref{table:wideresults}.

The results for LP in the top half of Table \ref{table:wideresults} are more interesting than those for MILP, and differ substantially from the corresponding results reported in Table \ref{table:mainresults}. Whereas option premia in excess of 0.1\% of the unit market investment were observed in between 5.1\% and 2.8\% of months using the moderate strike range (depending on the scale of investment), with the wide strike range we observe premia of this size in between 59.1\% and 12.1\% of months. Thus LP identifies many more substantial stochastic arbitrage opportunities using the wide range of strikes. We nevertheless see that realized portfolio performance is worse than that of the market portfolio. The mean excess returns of the enhanced portfolios are 2--3.5 percentage points below the mean SPX return, though these differences are only borderline statistically significant, or not significant.

\begin{figure*}
	\caption{Layover portfolio payoffs with wide strike range}\label{fig:widepayoffs}
	\centering
	\subcaptionbox{Portfolios selected with LP: 2004--2021}{
		\centering
		\begin{tikzpicture}[scale=3.8]
			\draw[black, thick,<->] (0,-.6) -- (0,.6);
			\draw[black, thick,<->] (-.1,-.5) -- (1.5,-.5);
			\draw ({((.7+2)/3-.875)*1.5/(1.075-.875)},{-.5cm+0.3pt}) -- ({((.7+2)/3-.875)*1.5/(1.075-.875)},{-.5cm-1pt})
			node[anchor=north,font=\scriptsize] {$.7$};
			\draw ({((.85+2)/3-.875)*1.5/(1.075-.875)},{-.5cm+0.3pt}) -- ({((.85+2)/3-.875)*1.5/(1.075-.875)},{-.5cm-1pt})
			node[anchor=north,font=\scriptsize] {$.85$};
			\draw ({((1+2)/3-.875)*1.5/(1.075-.875)},{-.5cm+0.3pt}) -- ({((1+2)/3-.875)*1.5/(1.075-.875)},{-.5cm-1pt})
			node[anchor=north,font=\scriptsize] {$1$};
			\draw ({((1.15+2)/3-.875)*1.5/(1.075-.875)},{-.5cm+0.3pt}) -- ({((1.15+2)/3-.875)*1.5/(1.075-.875)},{-.5cm-1pt})
			node[anchor=north,font=\scriptsize] {$1.15$};
			\draw (0.3pt,{(-.3+.1)*.6/.14*.5}) -- (-0.8pt,{(-.3+.1)*.6/.14*.5});
			\draw (0.3pt,{(-.2+.1)*.6/.14*.5}) -- (-0.8pt,{(-.2+.1)*.6/.14*.5});
			\draw (0.3pt,{(-.1+.1)*.6/.14*.5}) -- (-0.8pt,{(-.1+.1)*.6/.14*.5});
			\draw (0.3pt,{(0+.1)*.6/.14*.5}) -- (-0.8pt,{(0+.1)*.6/.14*.5});
			\draw (0.3pt,{(.1+.1)*.6/.14*.5}) -- (-0.8pt,{(.1+.1)*.6/.14*.5});
			\node[left,font=\scriptsize] at (0,{(-.3+.1)*.6/.14*.5}) {$-.3$};
			\node[left,font=\scriptsize] at (0,{(-.2+.1)*.6/.14*.5}) {$-.2$};
			\node[left,font=\scriptsize] at (0,{(-.1+.1)*.6/.14*.5}) {$-.1$};
			\node[left,font=\scriptsize] at (0,{(0+.1)*.6/.14*.5}) {$0$};
			\node[left,font=\scriptsize] at (0,{(.1+.1)*.6/.14*.5}) {$.1$};
			\draw[thin] (0,{(0+.1)*.6/.14*.5}) -- (1.5,{(0+.1)*.6/.14*.5});
			\draw[thick,only marks,olive] plot[mark=*,mark size=.15pt,xshift={0},xscale={1},yshift={.1*.6/.14*.5cm},yscale={.5}] file {realizedpayoffLPwide.txt};
			\node[font=\footnotesize] at ({3/4},-.725) {SPX at expiry};
			\node[rotate=90,font=\footnotesize] at (-.275,0) {Layover portfolio payoff};
		\end{tikzpicture}
		\hspace{1cm}
	}
	\subcaptionbox{Portfolios selected with LP: 2004--2012}{
		\centering
		\begin{tikzpicture}[scale=3.8]
			\draw[black, thick,<->] (0,-.6) -- (0,.6);
			\draw[black, thick,<->] (-.1,-.5) -- (1.5,-.5);
			\draw ({((.7+2)/3-.875)*1.5/(1.075-.875)},{-.5cm+0.3pt}) -- ({((.7+2)/3-.875)*1.5/(1.075-.875)},{-.5cm-1pt})
			node[anchor=north,font=\scriptsize] {$.7$};
			\draw ({((.85+2)/3-.875)*1.5/(1.075-.875)},{-.5cm+0.3pt}) -- ({((.85+2)/3-.875)*1.5/(1.075-.875)},{-.5cm-1pt})
			node[anchor=north,font=\scriptsize] {$.85$};
			\draw ({((1+2)/3-.875)*1.5/(1.075-.875)},{-.5cm+0.3pt}) -- ({((1+2)/3-.875)*1.5/(1.075-.875)},{-.5cm-1pt})
			node[anchor=north,font=\scriptsize] {$1$};
			\draw ({((1.15+2)/3-.875)*1.5/(1.075-.875)},{-.5cm+0.3pt}) -- ({((1.15+2)/3-.875)*1.5/(1.075-.875)},{-.5cm-1pt})
			node[anchor=north,font=\scriptsize] {$1.15$};
			\draw (0.3pt,{(-.3+.1)*.6/.14*.5}) -- (-0.8pt,{(-.3+.1)*.6/.14*.5});
			\draw (0.3pt,{(-.2+.1)*.6/.14*.5}) -- (-0.8pt,{(-.2+.1)*.6/.14*.5});
			\draw (0.3pt,{(-.1+.1)*.6/.14*.5}) -- (-0.8pt,{(-.1+.1)*.6/.14*.5});
			\draw (0.3pt,{(0+.1)*.6/.14*.5}) -- (-0.8pt,{(0+.1)*.6/.14*.5});
			\draw (0.3pt,{(.1+.1)*.6/.14*.5}) -- (-0.8pt,{(.1+.1)*.6/.14*.5});
			\node[left,font=\scriptsize] at (0,{(-.3+.1)*.6/.14*.5}) {$-.3$};
			\node[left,font=\scriptsize] at (0,{(-.2+.1)*.6/.14*.5}) {$-.2$};
			\node[left,font=\scriptsize] at (0,{(-.1+.1)*.6/.14*.5}) {$-.1$};
			\node[left,font=\scriptsize] at (0,{(0+.1)*.6/.14*.5}) {$0$};
			\node[left,font=\scriptsize] at (0,{(.1+.1)*.6/.14*.5}) {$.1$};
			\draw[thin] (0,{(0+.1)*.6/.14*.5}) -- (1.5,{(0+.1)*.6/.14*.5});
			\draw[thick,only marks,olive] plot[mark=*,mark size=.15pt,xshift={0},xscale={1},yshift={.1*.6/.14*.5cm},yscale={.5}] file {realizedpayoffearly.txt};
			\node[font=\footnotesize] at ({3/4},-.725) {SPX at expiry};
			\node[rotate=90,font=\footnotesize] at (-.275,0) {Layover portfolio payoff};
		\end{tikzpicture}
		\hspace{1cm}
	}
	\floatfoot{\emph{Notes:} Panel (a) displays the realized payoffs of the 215 layover portfolios selected using LP when $S=1$ and the SGT distribution is used to specify state probabilities. Panel (b) displays only the 108 payoffs realized during the first half of the sample period, i.e.\ 2004-2012. The horizontal axes measure the SPX on the date of expiry as a proportion of the SPX on the portfolio formation date. The range of included strikes is from 30\% below to 15\% above the current SPX.}
\end{figure*}

Direct examination of the payoffs delivered by the portfolios selected with LP provides insight into their poor realized performance. In Figure \ref{fig:widepayoffs}, which may be compared to Figure \ref{fig:Normalpayoffs} above, we plot the realized layover portfolio payoffs (with $S=1$) against the payoff of the market portfolio. Panel (a) displays all 215 payoffs in our full sample period 2004--2021, while panel (b) displays only the 108 payoffs for 2004--2012. A clear pattern is evident in both panels: layover payoffs are generally nonnegative and relatively small when the SPX falls or only modestly rises, but may be large and negative when the SPX return is very high. The choice of portfolios thus resembles a bet against the possibility of a large and positive market return. The bet pays off rather well during 2004--2012. In this subperiod the annualized mean excess portfolio return is an impressive 11.5\%. However, a series of unsuccessful bets during 2013--2021 brings the mean excess portfolio return down to 7.84\% for the full sample period, compared to the mean excess SPX return of 9.81\%.

The unsuccessful bet against large and positive SPX returns reflects a characteristic of the PIT histogram displayed in panel (b) of Figure \ref{fig:PDFs}. Here we see that our specification of state probabilities substantially underpredicts the top decile of SPX returns, with 13.6\% of 4498 daily observations of overlapping monthly returns exceeding the predicted top decile. The problem worsens at higher quantiles, with 7.5\% of observations exceeding the predicted top 5\%. Due to this apparent misspecification of top return probabilities, our portfolio selection procedure generates bets against large and positive SPX returns which are lost more often than expected. We are left with a cautionary tale about the pursuit of stochastic arbitrage with far out-of-the-money options. Tail probabilities for return distributions are inherently difficult to specify because, by definition, extreme returns are rarely observed. If tail probabilities are misspecified then our portfolio selection procedure will under- or overvalue far out-of-the-money options, guiding us to take positions in these options whose risk is misunderstood.

The results reported in the top half of Table \ref{table:wideresults} were obtained after eliminating 37 contracts from our option dataset which provided pure arbitrage opportunities, these having been identified by applying the methods introduced in \cite{CarrMadan05} as discussed in Section \ref{sec:data}. While pure arbitrage opportunities represent a special kind of stochastic arbitrage, they are not the focus of the present article and so we have excluded them to focus only on stochastic arbitrage opportunities which are not pure arbitrage opportunities. Nevetheless, in the bottom half of Table \ref{table:wideresults} we report the results which were obtained without eliminating these 37 contracts. The purpose is to reconcile our results with those reported in \cite{PostLongarela21}.

We see in the bottom half of Table \ref{table:wideresults} that the inclusion of pure arbitrage opportunities leads the enhanced portfolios selected with either LP or MILP to achieve an enormous annualized mean excess return of more than 20\% when $S=1$. The standard deviation of excess returns is also very large, driven almost entirely by the presence of a handful of extreme positive returns generated by pure arbitrage. When $S$ is increased to larger values these features of the enhanced portfolio returns disappear, and we see results much more similar to those reported in the top half of Table \ref{table:wideresults}. The reason for this sudden change when $S$ is increased is that the pure arbitrage opportunities present in our data are provided by only a small number of listed contracts. For a small investor (i.e., small $S$) the money to be made by exploiting these opportunities is large relative to the size of their overall investment, thus generating a large rate of return. For larger investors the amount of money to be made is the same, but is much smaller relative to the size of their overall investment, so the effect on the rate of return is greatly reduced.

The results we report for LP in the bottom half of Table \ref{table:wideresults} may be compared to those reported in Table 2 in \cite{PostLongarela21}. The empirical analysis undertaken there differs from what has been done here in three main respects: the sample period is 2004--2018, the market risk premium and variance risk premium are chosen differently, and state probabilities are specified using a binomial approximation to the normal distribution rather than using the SGT distribution. It is found there, as has been found here, that with $S=1$ the excess portfolio return has a very large mean, standard deviation and positive skew, whereas for larger values of $S$ these features are not present and option portfolios perform poorly. Our results strongly suggest that the large portfolio returns reported in \cite{PostLongarela21} for $S=1$ may be driven by pure arbitrage. The authors do not claim to exclude pure arbitrage, and emphasize at several points that the large monthly portfolio returns reported for $S=1$ should not be viewed as a robust finding.

\section{Concluding remarks}\label{sec:conclusion}

The central empirical finding of this article is that stochastic arbitrage opportunities are not obviously apparent in the market for SPX options. Option prices at moderate strikes appear to be broadly consistent with a simple specification of the SPX return distribution obtained by shifting and scaling a skewed generalized t-distribution. At more extreme strikes, option mispricing is difficult to assess due to the relative paucity of outlying return observations. If state probabilities are misjudged---as is likely to be the case when far out-of-the-money options are under consideration---then fairly priced options may appear to be mispriced, and trading strategies executed which are expected to generate stochastic arbitrage but do not in fact do so.

Our negative findings are consistent with the empirical results reported in \cite{PostLongarela21} but contrast with those reported in \cite{ConstantinidesCzerwonkoPerrakis20}. The latter article finds evidence of stochastic arbitrage opportunities with SPX options during the years 1990--2013. Aside from the different sample periods, an important way in which the analyses undertaken here and in \cite{PostLongarela21} differ from the analysis undertaken in \cite{ConstantinidesCzerwonkoPerrakis20} is in our requirement that the quantities of options bought and written cannot exceed the quoted bid and ask sizes. This constraint prevents us from generating large profits from mispriced options listed only in small quantities, particularly when the scale of investment is larger.

A potentially undesirable aspect of the option portfolio selection algorithms proposed here and in \cite{PostLongarela21} is the fact that there is no limit on the number of nonzero positions taken. A selected portfolio may include a mixture of sensible positions in mispriced options, and less sensible positions driven by the misspecification of state probabilities. Ideally one would like to impose a sparsity condition on portfolio weights in such a way that nonzero positions are taken only in those options which present a clear opportunity for stochastic arbitrage. Limiting the number of nonzero option positions would also simplify the execution of trades. We leave further exploration of this matter to future research.

\section*{Acknowledgments}
	
The authors thank the Editor (Geert Bekaert) and two anonymous referees for valuable comments and suggestions. The authors further thank Xinyi Luo for research assistance, James Luedtke for advice on mixed-integer linear programming, and Stylianos Perrakis and Thierry Post for helpful comments on a draft manuscript. Special thanks are due to Lawrence Schmidt, whose research on this general topic with Beare at the University of California San Diego during 2012--15 provided foundations for the present study.
	
Beare acknowledges financial support from the National University of Singapore through the Isaac Manasseh Meyer Fellowship. Seo acknowledges financial support from the Singapore Ministry of Education under its Academic Research Fund Tier 1 (Project ID: R-122-000-297-115).

\appendix
\setcounter{equation}{0}
\renewcommand\theequation{\Alph{section}.\arabic{equation}}
\setcounter{table}{0}
\renewcommand\thetable{\Alph{section}.\arabic{table}}
\setcounter{figure}{0}
\renewcommand\thefigure{\Alph{section}.\arabic{figure}}

\section{Computational efficiency of LP}\label{sec:LPstar}

\begin{table*}[b!]
	\centering
	\begin{threeparttable}
		\caption{Relative computational efficiency of LP and LP*}\label{table:SA2runtime}
		\begin{tabular}{lllrrrrrrrr}
			\toprule
			\multicolumn{3}{c}{}&\multicolumn{2}{c}{2004--21 average}&&\multicolumn{2}{c}{2021 average}\\\cmidrule{4-5}\cmidrule{7-8}
			Strike range&Optimizer&Program&Input file (MB)&Runtime (s)&&Input file (MB)&Runtime (s)\\
			\midrule
			\multirow{2}{*}{Moderate}&\multirow{2}{*}{Gurobi}&LP&1.2&0.04&&5.1&0.19\\
			&&LP*&35.2&0.34&&219.1&2.76\\
			\midrule
			\multirow{2}{*}{Moderate}&\multirow{2}{*}{CPLEX}&LP&1.2&0.05&&5.1&0.28\\
			&&LP*&35.2&0.24&&219.1&1.90\\
			\midrule
			\multirow{2}{*}{Wide}&\multirow{2}{*}{Gurobi}&LP&10.4&0.79&&44.7&4.36\\
			&&LP*&811.4&25.41&&4874.9&243.92\\
			\midrule
			\multirow{2}{*}{Wide}&\multirow{2}{*}{CPLEX}&LP&10.4&0.65&&44.7&3.30\\
			&&LP*&811.4&45.83&&4874.9&709.08\\
			\bottomrule
		\end{tabular}
	\end{threeparttable}
	\floatfoot{\emph{Notes:} Computations were performed using Gurobi Optimizer 11.0.3 and CPLEX Optimizer 22.1.1 on a machine with a 2 GHz processor and 128 GB of memory.}
\end{table*}

As discussed in Section \ref{sec:sd2}, the linear program LP differs from the linear program proposed in \citet{PostLongarela21} while providing an identical solution for the layover portfolio. The formulation of the latter linear program relies on the fact that a layover portfolio $(\boldsymbol{\alpha},\boldsymbol{\beta})\in\mathcal P$ satisfies the second-order stochastic dominance constraint \eqref{eq:sdconstraint} if and only if there exists an $n\times n$ nonnegative matrix $\boldsymbol{\Psi}$ such that
\begin{align}
	\boldsymbol{\Psi}\boldsymbol{\mu}&\leq\mathbf{T}\boldsymbol{\mu},\label{eq:PLconstraint1}\\
	-\boldsymbol{\Psi}^\top-\boldsymbol{\Theta}^\top(\boldsymbol{\alpha}-\boldsymbol{\beta})\mathbf{1}_n^\top&\leq\mathbf{x}\mathbf{1}_n^\top-\mathbf{1}_n\mathbf{x}^\top,\label{eq:PLconstraint2}
\end{align}
where $\mathbf{T}$ is the $n\times n$ strictly lower triangular matrix with $(j,k)$th entry equal to $x_j-x_k$ for $j>k$. This is a consequence of results established in \citet{DentchevaRuszczynski03} and \citet{Kuosmanen04}; see also closely related results in \citet{RockafellarUryasev00} and \citet{Post03}, and the discussion following Lemma 2.1 in \citet{Luedtke08}. It justifies using the following linear program proposed in \citet{PostLongarela21} to compute the optimal layover portfolio.
\begin{LP*}
	Choose two $m\times1$ nonnegative vectors $\boldsymbol{\alpha},\boldsymbol{\beta}$ and an $n\times n$ nonnegative matrix $\boldsymbol{\Psi}$ to maximize $-\mathbf{p}^\top\boldsymbol{\alpha}+\mathbf{q}^\top\boldsymbol{\beta}$ subject to the constraints \eqref{eq:PLconstraint1}, \eqref{eq:PLconstraint2}, and \eqref{eq:PLconstraint3}.
\end{LP*}
While LP and LP* provide identical solutions for $\boldsymbol{\alpha}$ and $\boldsymbol{\beta}$, the former program is computationally advantageous. The reason is that both sides of the inequality in \eqref{eq:PLconstraint2} are $n\times n$ matrices, meaning that this inequality is comprised of $n^2$ scalar inequalities. Constraints \eqref{eq:PLconstraint1} and \eqref{eq:PLconstraint2} in LP* therefore comprise a total of $n^2+n$ scalar inequalities, an order of magnitude greater than the $4n$ scalar equalities and inequalities comprising constraints \eqref{eq:Lconstraint1}, \eqref{eq:Lconstraint2}, \eqref{eq:Lconstraint3} and \eqref{eq:Lconstraint4} in LP.

We found in our empirical application that the gap in computational efficiency between LP and LP* is large. Table \ref{table:SA2runtime} reports a comparison of the the two programs when applied to the 215 portfolio choice problems in our empirical application with the unit market depth constraint and with state probabilities specified as described in Section \ref{sec:phys}. Two leading optimizers were used: Gurobi Optimizer 11.0.3 and CPLEX Optimizer 22.1.1. Input files for Gurobi and CPLEX in the MPS (Mathematical Programming System) format were created using the JuMP (Julia Mathematical Programming) package for the Julia programming language. Runtimes and input file sizes were larger toward the end of our sample period due to the fact that growth in the SPX over time necessitates the use of more atoms spread at 5 index point increments to cover the range of strikes. The average input file size for the final year in our sample period was 219.1 MB for LP* and 5.1 MB for LP using the moderate strike range, and was 4874.9 MB for LP* and 44.7 MB for LP using the wide strike range. Gurobi took an average of 2.76 seconds to solve LP* in the final year with the moderate strike range, compared to only 0.19 seconds for LP. With the wide strike range, Gurobi took an average of 243.92 seconds to solve LP* in the final year, compared to only 4.36 seconds for LP. Runtimes for Gurobi and CPLEX were generally similar, except when solving LP* using the wide strike range. See \citet{Luedtke08} for more extensive results documenting the gains in computational efficiency achieved by imposing second-order stochastic dominance using constraints \eqref{eq:Lconstraint1}, \eqref{eq:Lconstraint2}, \eqref{eq:Lconstraint3} and \eqref{eq:Lconstraint4} rather than constraints \eqref{eq:PLconstraint1} and \eqref{eq:PLconstraint2}.

\section{Starting values for MILP}\label{sec:MILP}
\setcounter{equation}{0}
\setcounter{table}{0}
\setcounter{figure}{0}

\begin{table*}[b!]
	\centering
	\begin{threeparttable}
		\caption{Evaluation of MILP-ST starting values}\label{table:SA1runtime}
		\begin{tabular}{rrrrr}
			\toprule
			Optimizer&Starting values&Max.\ runtime (s)&Pctg.\ premia $> 0.1$\%&Avg.\ premium (\%)\\
			\midrule
			Gurobi&Default&10&18.6&0.112\\
			Gurobi&Default&300&24.7&0.164\\
			Gurobi&MILP-ST&9&24.2&0.158\\
			\midrule
			CPLEX&Default&10&20.0&0.116\\
			CPLEX&Default&300&22.3&0.155\\
			CPLEX&MILP-ST&9&23.7&0.155\\
			\midrule
			\multicolumn{3}{r}{MILP-ST without further optimization}&22.3&0.152\\
			\bottomrule
		\end{tabular}
	\end{threeparttable}
	\floatfoot{\emph{Notes:} Computations were performed using Gurobi Optimizer 11.0.3 and CPLEX Optimizer 22.1.1 on a machine with a 3.5 GHz processor and 128 GB of memory.}
\end{table*}

The portfolios chosen to solve MILP in Section \ref{sec:empirical} were computed by supplying Gurobi with starting values obtained using the following algorithm.	
\begin{MILP-ST}\leavevmode
	\begin{enumerate}[wide, labelwidth=!, labelindent=0pt, label=\textbf{\arabic*}.]
		\item Set $\boldsymbol{\alpha}^0\coloneqq\mathbf{0}_m$, $\boldsymbol{\beta}^0\coloneqq\mathbf{0}_m$, $\boldsymbol{\xi}^0\coloneqq\boldsymbol{\mu}$ and $\boldsymbol{\Psi}^0\coloneqq\mathbf{I}_n$. Choose $\boldsymbol{\alpha}^1$ and $\boldsymbol{\beta}^1$ to solve LP. If $-\mathbf{p}^\top\boldsymbol{\alpha}^1+\mathbf{q}^\top\boldsymbol{\beta}^1=0$ then $\boldsymbol{\alpha}^0$, $\boldsymbol{\beta}^0$, $\boldsymbol{\xi}^0$ and $\boldsymbol{\Psi}^0$ solve MILP. Otherwise, set $\pi\coloneqq0$ and $j\coloneqq1$.
		\item Set $k\coloneqq1$ and $\boldsymbol{\omega}^j\coloneqq\mathbf{x}+\boldsymbol{\Theta}^\top(\boldsymbol{\alpha}^j-\boldsymbol{\beta}^j)$. Sort $\boldsymbol{\omega}^j$ to obtain distinct integers $i_1,\dots,i_n$ such that $\boldsymbol{\omega}^j_{i_1}\leq\boldsymbol{\omega}^j_{i_2}\leq\cdots\leq\boldsymbol{\omega}^j_{i_n}$. Construct an $n\times n$ binary matrix $\boldsymbol{\Psi}^j$ by first setting $\boldsymbol{\Psi}^j\coloneqq\mathbf{0}_{n\times n}$ and then applying the following algorithm:\\
		\hspace*{.5em}\begin{tabular}[t]{l}
			\textbf{for } $t\coloneqq1$ \textbf{ to } $n$ \textbf{ do}\\
			\vline \hspace*{1em} \textbf{while } $k\leq n$ \textbf{ and } $\sum_{s=1}^k\boldsymbol{\mu}_{i_s}\leq\sum_{s=1}^t\boldsymbol{\mu}_s$ \textbf{ do}\\
			\vline \hspace*{1.3em} \vline \hspace*{1em} $\boldsymbol{\Psi}^j_{i_k,t}\coloneqq1$;\\
			\vline \hspace*{1.3em} \vline \hspace*{2.3em} $k\coloneqq k+1$;\\
			\vline \hspace*{1em} \textbf{end}\\
			\textbf{end}
		\end{tabular}\\
		Set $\boldsymbol{\xi}^j\coloneqq(\boldsymbol{\Psi}^j)^\top\boldsymbol{\mu}$.
		\item Choose nonnegative $m\times 1$ vectors $\boldsymbol{\alpha}^{j+1}$ and $\boldsymbol{\beta}^{j+1}$ to maximize $-\mathbf{p}^\top\boldsymbol{\alpha}^{j+1}+\mathbf{q}^\top\boldsymbol{\beta}^{j+1}$ subject to the constraints $\boldsymbol{\Psi}^j\mathbf{x}-\boldsymbol{\Theta}^\top(\boldsymbol{\alpha}^{j+1}-\boldsymbol{\beta}^{j+1})\leq\mathbf{x}$ and $\mathbf{A}\boldsymbol{\alpha}^{j+1}+\mathbf{B}\boldsymbol{\beta}^{j+1}\leq\mathbf{c}$. If there is no feasible solution or if the maximum attained does not exceed $\pi$ then go to step 4. Otherwise, set $\pi\coloneqq-\mathbf{p}^\top\boldsymbol{\alpha}^{j+1}+\mathbf{q}^\top\boldsymbol{\beta}^{j+1}$, then set $j\coloneqq j+1$ and return to step 2.
		\item If $j=1$ then we use $\boldsymbol{\alpha}^{0}$, $\boldsymbol{\beta}^{0}$, $\boldsymbol{\xi}^{0}$ and $\boldsymbol{\Psi}^{0}$ as starting values for MILP. If $j>1$ then we use $\boldsymbol{\alpha}^{j}$, $\boldsymbol{\beta}^{j}$, $\boldsymbol{\xi}^{j-1}$ and $\boldsymbol{\Psi}^{j-1}$ as starting values for MILP.
	\end{enumerate}
\end{MILP-ST}

Step 1 of MILP-ST involves solving LP, the linear relaxation of MILP. This can be achieved in well under one second (see Table \ref{table:SA2runtime}). Steps 2 and 3 may be iterated any number of times, but we found in our empirical implementation that the number of iterations never exceeded 16. Each iteration of step 3 requires solving a linear program, but this program is much smaller than LP as it contains only $2m$ choice variables, compared to $2m+n+n^2$ choice variables in LP. The total runtime for MILP-ST was always less than one second in our empirical implementations with the moderate strike range.

MILP-ST is an adaptation of Algorithm 1 in \citet{Luedtke08}. It is proposed there as a customized heuristic to be called repeatedly during optimization, but here we use it only for computing starting values. The key innovation is the construction of $\boldsymbol{\Psi}^j$ in step 2. As discussed in \citet{Luedtke08}, $\boldsymbol{\Psi}^j$ and $\boldsymbol{\xi}^j$ are guaranteed to satisfy constraints \eqref{eq:Lconstraint1}, \eqref{eq:Lconstraint2}, \eqref{eq:Lconstraint3} in MILP. Therefore, if $\boldsymbol{\alpha}^{j+1}$ and $\boldsymbol{\beta}^{j+1}$ solve the linear program in step 3, then combining $\boldsymbol{\Psi}^j$ and $\boldsymbol{\xi}^j$ with $\boldsymbol{\alpha}^{j+1}$ and $\boldsymbol{\beta}^{j+1}$ yields a feasible solution to MILP. The hope is that the sorting procedure used to construct successive feasible solutions leads us toward good starting values for MILP.

Results reported in Table \ref{table:SA1runtime} support the use of MILP-ST for choosing starting values. For the 215 portfolio choice problems in our sample with the normal specification of state probabilities and unit market depth constraint, we measured the effectiveness of solution procedures for MILP in terms of the percentage of months in which they identified a portfolio with option premium exceeding 0.1\% of the market investment, and in terms of the average premium identified. Results for the Gurobi and CPLEX optimizers were similar. Both optimizers produced better solutions with a maximum runtime of 9 seconds and the MILP-ST starting values than they did with a maximum runtime of 10 seconds and the default starting values (i.e., starting values corresponding to zero positions in options). Increasing the maximum runtime to 300 seconds with the default starting values yielded similar performance to that obtained with the MILP-ST starting values and a maximum runtime of 9 seconds.

The final row of Table \ref{table:SA1runtime} reports the average performance of the MILP-ST starting values if they are used to directly construct option portfolios without further optimization. It is apparent that MILP-ST identifies a good feasible solution to MILP. The improvement to the MILP-ST starting values produced by an additional 9 seconds of runtime with Gurobi or CPLEX is modest. If runtime is a significant concern, MILP-ST can be used to directly compute good feasible solutions to MILP in a fraction of a second.

\section{Results with time-varying risk premia or asymmetry}\label{sec:tv}

The specification of state probabilities described in Section \ref{sec:phys} is based on a time-invariant multiplicative variance risk premium of $1.41^2$, time-invariant market risk premium of $6.67\%$, and time-invariant parametrization of the SGT distribution. In this appendix we investigate how our empirical results are affected when time-varying specifications of risk premia or asymmetry are employed. For brevity we confine attention to how the key results reported in the first of the four panels in each of Tables \ref{table:mainresults} and \ref{table:wideresults} are affected. These results pertain to option portfolios selected with LP using either the moderate or wide strike range, with pure arbitrage opportunities excluded.

\subsection{Time-varying multiplicative variance risk premium}\label{sec:mvrptv}

We introduce time-variation in the multiplicative variance risk premium following the approach described in \citet[p.~309]{ProkopczukSimen14}. For each trading day $t$, let $\mathrm{IV}_{t,\,t+21}$ be the one-month-ahead option implied variance for the SPX return (as determined by the VIX, and assuming 21 trading days in each month), and let $\mathrm{RV}_{t,\,t+21}$ be the sum of the daily realized variances (computed from 5 minute SPX return data as described in Section \ref{sec:data}) for trading days $t+1$ through $t+21$. Following \citet{ProkopczukSimen14}, we define the one-month-ahead multiplicative variance risk premium for trading day $t$ to be
\begin{equation}\label{eq:MVRP}
	\mathrm{MVRP}_{t,\,t+21}=\frac{1}{252-21}\sum_{s=t-251}^{t-21}\frac{\mathrm{IV}_{s,\,s+21}}{\mathrm{RV}_{s,\,s+21}}.
\end{equation}
Note that $\mathrm{MVRP}_{t,\,t+21}$ is computed from data observed on or before trading day $t$. Figure \ref{fig:mvrptv} displays a plot of the computed time-varying multiplicative variance risk premium.

\begin{figure}[b!]
	\caption{Time-varying multiplicative variance risk premium}\label{fig:mvrptv}
	\centering
	\begin{tikzpicture}[scale=3.5]
		\draw[black, thick,->] (0,0) -- (0,1.1);
		\draw[black, thick] (0,0) -- (1.3,0);
		\draw (0,0.3pt) -- (0,-0.8pt);
		\node[anchor=north,font=\scriptsize] at (0,-0.8pt) {$2004$};
		\draw ({(4*365+1)*1.3/6575},0.3pt) -- ({(4*365+1)*1.3/6575},-0.8pt);
		\node[anchor=north,font=\scriptsize] at ({(4*365+1)*1.3/6575},-0.8pt) {$2008$};
		\draw ({(8*365+2)*1.3/6575},0.3pt) -- ({(8*365+2)*1.3/6575},-0.8pt);
		\node[anchor=north,font=\scriptsize] at ({(8*365+2)*1.3/6575},-0.8pt) {$2012$};
		\draw ({(12*365+3)*1.3/6575},0.3pt) -- ({(12*365+3)*1.3/6575},-0.8pt);
		\node[anchor=north,font=\scriptsize] at ({(12*365+3)*1.3/6575},-0.8pt) {$2016$};
		\draw ({(16*365+4)*1.3/6575},0.3pt) -- ({(16*365+4)*1.3/6575},-0.8pt);
		\node[anchor=north,font=\scriptsize] at ({(16*365+4)*1.3/6575},-0.8pt) {$2020$};
		\draw (0.3pt,0) -- (-0.8pt,0);
		\draw (0.3pt,{1/4}) -- (-0.8pt,{1/4});
		\draw (0.3pt,{2/4}) -- (-0.8pt,{2/4});
		\draw (0.3pt,{3/4}) -- (-0.8pt,{3/4});
		\draw (0.3pt,1) -- (-0.8pt,1);
		\node[left,font=\scriptsize] at (0,0) {$1$};
		\node[left,font=\scriptsize] at (0,{1/4}) {$1.2$};
		\node[left,font=\scriptsize] at (0,{2/4}) {$1.4$};
		\node[left,font=\scriptsize] at (0,{3/4}) {$1.6$};
		\node[left,font=\scriptsize] at (0,1) {$1.8$};
		\draw[blue,thick] plot file {mvrp.txt};
		\draw[dashed] (0,{(1.4076-1)/.8}) -- (1.3,{(1.4076-1)/.8});
		\node[font=\footnotesize] at (.6,-.225) {Year};
		\node[rotate=90,font=\footnotesize] at (-.275,.5) {$\sqrt{\mathrm{MVRP}_{t,\,t+21}}$};
	\end{tikzpicture}
	\floatfoot{\emph{Notes:} $\mathrm{MVRP}_{t,\,t+21}$ is computed for each trading day $t$ as in Equation \eqref{eq:MVRP}. See also Equation (7) in \citet[p.~309]{ProkopczukSimen14}. The dashed line corresponds to the time-invariant multiplicative variance risk premium used in Sections \ref{sec:results}--\ref{sec:resultswide}.}
\end{figure}

We repeated the analysis reported in the first of the four panels in each of Tables \ref{table:mainresults} and \ref{table:wideresults} using $\mathrm{MVRP}_{t,\,t+21}$ in place of the time-invariant multiplicative variance risk premium. Table \ref{table:mvrptv} presents key aspects of the updated results. It is apparent that using the time-varying multiplicative variance risk premium leads to only minor changes. Our central findings---that substantial stochastic arbitrage opportunities are rarely identified with the moderate strike range, more frequently identified with the wide strike range, and in either case do not generate realized profits---are unchanged.

\begin{table*}[t!]
	\centering
	{\fontsize{9}{10.8}\selectfont
		\begin{threeparttable}
			\caption{Results with time-varying multiplicative variance risk premium}\label{table:mvrptv}
			\begin{tabular}{clccccc}
				\toprule
				&&&\multicolumn{4}{c}{Enhanced portfolio}\\\cmidrule{4-7}
				Strike range&& SPX & $S=1$ & $S=10$ & $S=100$ & $S=1000$\\
				\midrule
				\multirow{4}{*}{Moderate}&Pctg.\ premia $>0.1\% $ mkt.\ investment&&  5.6 &  4.7 &  4.2 &  3.7 \\
				&Realized mean excess return&9.81& 9.01 & 8.94 & 9.00 & 9.25 \\
				&Realized Sortino ratio&0.71& 0.66 & 0.65 & 0.65 & 0.67 \\
				&Equal mean excess return p-value&& 0.301 & 0.248 & 0.265 & 0.356 \\
				\midrule
				\multirow{4}{*}{Wide}&Pctg.\ premia $>0.1\% $ mkt.\ investment&& 58.6 & 48.4 & 32.1 & 13.0 \\
				&Realized mean excess return&9.81& 8.20 & 6.20 & 6.21 & 7.89 \\
				&Realized Sortino ratio&0.71& 0.59 & 0.45 & 0.45 & 0.58 \\
				&Equal mean excess return p-value&& 0.577 & 0.170 & 0.084 & 0.140 \\
				\bottomrule
			\end{tabular}
		\end{threeparttable}
	}
	\floatfoot{\emph{Notes:} The table reports characteristics of the 215 portfolios selected with LP. State probabilities are specified as described in Section \ref{sec:phys}, but using the time-varying multiplicative variance risk premium plotted in Figure \ref{fig:mvrptv}. Option contracts presenting a pure arbitrage opportunity are excluded. The top half of the table may be compared to the top quarter of Table \ref{table:mainresults}, while the bottom half of the table may be compared to the top quarter of Table \ref{table:wideresults}.}
\end{table*}

\subsection{Time-varying market risk premium}\label{sec:mrptv}

Specifying a time-varying market risk premium is inherently difficult due to the unpredictability of market returns. Basing portfolio choices on a weak predictor of market returns may do more harm than good, particularly if the precise nature of the relationship between the predictor and the market return is not well understood. To provide an example of the difficulties which may arise we consider a time-varying specification of the market risk premium which varies linearly with an estimate of the difference between the implied and realized variance of the future market return. Specifically, for each trading day $t$ we compute the estimated additive variance risk premium
\[\mathrm{AVRP}_{t,\,t+21}=\mathrm{IV}_{t,\,t+21}-\mathrm{RV}_{t-21,\,t},\]
where $\mathrm{IV}_{t,\,t+21}$ and $\mathrm{RV}_{t-21,\,t}$ are defined as in Section \ref{sec:mvrptv}. Note that $\mathrm{AVRP}_{t,\,t+21}$ is computed from data observed on or before trading day $t$. Based on an analysis of data from the years 1990--2007, \citet{BollerslevTauchenZhou09} argue that the additive variance risk premium may be useful for predicting future market returns at a quarterly horizon. Evidence for predictability at a monthly horizon is mixed. \citet[p.~4480]{BollerslevTauchenZhou09} report the estimated linear regression equation
\begin{equation}\label{eq:BTZ09}
	Y_{t,t+21}=-0.55+0.39\times\mathrm{AVRP}_{t,\,t+21}+\text{residual},
\end{equation}
where $Y_{t,\,t+21}$ is the one-month SPX excess return realized on trading day $t+21$. Equation \eqref{eq:BTZ09} indicates a positive relationship between the additive variance risk premium and one-month-ahead SPX excess return. However, the autocorrelation-robust t-statistic for the estimated slope coefficient is only 1.76, and the adjusted regression $R^2$ is only 1.07\%. Thus Equation \eqref{eq:BTZ09} provides, at best, a weak signal of the one-month-ahead SPX excess return. Further empirical analysis reported in \citet{BekaertHoerova14}, based on data from 1990--2010 and a modified estimate of the additive variance risk premium, provides somewhat stronger evidence of predictability at the monthly horizon.

Equation \eqref{eq:BTZ09} may be used to construct a time-varying market risk premium depending linearly on the additive variance risk premium. Figure \ref{fig:mrptv}(a) plots the time-varying market risk premium implied by Equation \eqref{eq:BTZ09} for our sample period. The market risk premium varies wildly over time, sometimes taking extremely large positive or negative values, and the mean premium is only 2.8\%. To obtain more plausible premia we shifted the estimated values upward to produce a mean of 6.67\%, and winsorized by restricting values to be between 0\% and 15\%. The resulting estimated premia are plotted in Figure \ref{fig:mrptv}(b).

\begin{figure*}[t!]
	\caption{Estimated time-varying market risk premium}\label{fig:mrptv}
	\centering
	\subcaptionbox{Raw estimates from Equation \eqref{eq:BTZ09}}{
		\centering
		\begin{tikzpicture}[scale=3.5]
			\draw[black, thick,->] (0,0) -- (0,1.1);
			\draw[black, thick] (0,0) -- (1.3,0);
			\draw (0,0.3pt) -- (0,-0.8pt);
			\node[anchor=north,font=\scriptsize] at (0,-0.8pt) {$2004$};
			\draw ({(4*365+1)*1.3/6575},0.3pt) -- ({(4*365+1)*1.3/6575},-0.8pt);
			\node[anchor=north,font=\scriptsize] at ({(4*365+1)*1.3/6575},-0.8pt) {$2008$};
			\draw ({(8*365+2)*1.3/6575},0.3pt) -- ({(8*365+2)*1.3/6575},-0.8pt);
			\node[anchor=north,font=\scriptsize] at ({(8*365+2)*1.3/6575},-0.8pt) {$2012$};
			\draw ({(12*365+3)*1.3/6575},0.3pt) -- ({(12*365+3)*1.3/6575},-0.8pt);
			\node[anchor=north,font=\scriptsize] at ({(12*365+3)*1.3/6575},-0.8pt) {$2016$};
			\draw ({(16*365+4)*1.3/6575},0.3pt) -- ({(16*365+4)*1.3/6575},-0.8pt);
			\node[anchor=north,font=\scriptsize] at ({(16*365+4)*1.3/6575},-0.8pt) {$2020$};
			\draw (0.3pt,0) -- (-0.8pt,0);
			\draw (0.3pt,{1/4}) -- (-0.8pt,{1/4});
			\draw (0.3pt,{2/4}) -- (-0.8pt,{2/4}); 
			\draw (0.3pt,{3/4}) -- (-0.8pt,{3/4});
			\draw (0.3pt,1) -- (-0.8pt,1);
			\node[left,font=\scriptsize] at (0,0) {$-0.9$};
			\node[left,font=\scriptsize] at (0,{1/4}) {$-0.6$};
			\node[left,font=\scriptsize] at (0,{2/4}) {$-0.3$};
			\node[left,font=\scriptsize] at (0,{3/4}) {$0$};
			\node[left,font=\scriptsize] at (0,1) {$0.3$};
			\draw[blue,thick] plot file {mktp.txt};
			\draw[dashed] (0,{(.0667+.9)/1.2}) -- (1.3,{(.0667+.9)/1.2});
			\node[font=\footnotesize] at (.6,-.225) {Year};
			\node[rotate=90,font=\footnotesize] at (-.275,.5) {Market risk premium};
		\end{tikzpicture}
		\hspace{.3cm}
	}
	\hspace{.4cm}
	\subcaptionbox{Recentered and winsorized estimates}{
		\centering
		\begin{tikzpicture}[scale=3.5]
			\draw[black, thick,->] (0,0) -- (0,1.1);
			\draw[black, thick] (0,0) -- (1.3,0);
			\draw (0,0.3pt) -- (0,-0.8pt);
			\node[anchor=north,font=\scriptsize] at (0,-0.8pt) {$2004$};
			\draw ({(4*365+1)*1.3/6575},0.3pt) -- ({(4*365+1)*1.3/6575},-0.8pt);
			\node[anchor=north,font=\scriptsize] at ({(4*365+1)*1.3/6575},-0.8pt) {$2008$};
			\draw ({(8*365+2)*1.3/6575},0.3pt) -- ({(8*365+2)*1.3/6575},-0.8pt);
			\node[anchor=north,font=\scriptsize] at ({(8*365+2)*1.3/6575},-0.8pt) {$2012$};
			\draw ({(12*365+3)*1.3/6575},0.3pt) -- ({(12*365+3)*1.3/6575},-0.8pt);
			\node[anchor=north,font=\scriptsize] at ({(12*365+3)*1.3/6575},-0.8pt) {$2016$};
			\draw ({(16*365+4)*1.3/6575},0.3pt) -- ({(16*365+4)*1.3/6575},-0.8pt);
			\node[anchor=north,font=\scriptsize] at ({(16*365+4)*1.3/6575},-0.8pt) {$2020$};
			\draw (0.3pt,0) -- (-0.8pt,0);
			\draw (0.3pt,{1/3}) -- (-0.8pt,{1/3});
			\draw (0.3pt,{2/3}) -- (-0.8pt,{2/3});
			\draw (0.3pt,1) -- (-0.8pt,1);
			\node[left,font=\scriptsize] at (0,0) {$0$};
			\node[left,font=\scriptsize] at (0,{1/3}) {$0.05$};
			\node[left,font=\scriptsize] at (0,{2/3}) {$0.1$};
			\node[left,font=\scriptsize] at (0,1) {$0.15$};
			\draw[blue,thick] plot file {mktp_reg.txt};
			\draw[dashed] (0,{.0667/.15}) -- (1.3,{.0667/.15});
			\node[font=\footnotesize] at (.6,-.225) {Year};
			\node[rotate=90,font=\footnotesize] at (-.275,.5) {Market risk premium};
		\end{tikzpicture}
		\hspace{.3cm}
	}
	\floatfoot{\emph{Notes:} Panel (a) displays the fitted value for $Y_{t,\,t+21}$ implied by the regression equation \eqref{eq:BTZ09} at each portfolio formation date $t$. Panel (b) displays the same values after shifting upward to obtain a mean value of 6.67\%, and then winsorizing at 0\% and 15\%. The dashed lines correspond to the time-invariant market risk premium of 6.67\% used in the analysis reported in Section \ref{sec:empirical}.}
\end{figure*}

We repeated the analysis reported in the first of the four panels in each of Tables \ref{table:mainresults} and \ref{table:wideresults} using the time-varying market risk premia plotted in panels (a) and (b) of Figure \ref{fig:mrptv} to specify state probabilities. The key findings are reported in Table \ref{table:mrptv}. With either choice of time-varying market risk premium the results differ greatly from those obtained with the fixed market premium. Plainly, a severe misspecification of state probabilities caused by excessive variation in the market risk premium leads the selected portfolios to perform extremely poorly.

\begin{table*}[b!]
	\centering
	{\fontsize{8.4}{10.8}\selectfont
		\begin{threeparttable}
			\caption{Results with time-varying market risk premium}\label{table:mrptv}
			\begin{tabular}{cclccccc}
				\toprule
				&&&&\multicolumn{4}{c}{Enhanced portfolio}\\\cmidrule{4-7}
				Strike range&MRP&& SPX & $S=1$ & $S=10$ & $S=100$ & $S=1000$\\
				\midrule
				\multirow{4}{*}{Moderate}&\multirow{4}{*}{Fig.\ \ref{fig:mrptv}(a)}&Pctg.\ premia $>0.1\% $ mkt.\ investment&& 65.1 & 59.1 & 47.4 & 31.6 \\
				&&Realized mean excess return&9.81& 3.62 & 3.30 & 4.25 & 7.20 \\
				&&Realized Sortino ratio&0.71& 0.27 & 0.25 & 0.32 & 0.54 \\
				&&Equal mean excess return p-value&& 0.032 & 0.016 & 0.014 & 0.069 \\
				\midrule
				\multirow{4}{*}{Moderate}&\multirow{4}{*}{Fig.\ \ref{fig:mrptv}(b)}&Pctg.\ premia $>0.1\% $ mkt.\ investment&& 45.6 & 42.3 & 37.7 & 25.1 \\
				&&Realized mean excess return&9.81& 3.97 & 4.39 & 6.01 & 7.91 \\
				&&Realized Sortino ratio&0.71& 0.30 & 0.33 & 0.44 & 0.59 \\
				&&Equal mean excess return p-value&& 0.011 & 0.010 & 0.045 & 0.124 \\
				\midrule
				\multirow{4}{*}{Wide}&\multirow{4}{*}{Fig.\ \ref{fig:mrptv}(a)}&Pctg.\ premia $>0.1\% $ mkt.\ investment&& 84.7 & 80.5 & 70.7 & 50.2 \\
				&&Realized mean excess return&9.81& -2.68 & -3.41 & -2.64 & -0.24 \\
				&&Realized Sortino ratio&0.71& -0.21 & -0.27 & -0.20 & -0.02 \\
				&&Equal mean excess return p-value&& 0.004 & 0.001 & 0.001 & 0.000 \\
				\midrule
				\multirow{4}{*}{Wide}&\multirow{4}{*}{Fig.\ \ref{fig:mrptv}(b)}&Pctg.\ premia $>0.1\% $ mkt.\ investment&& 63.7 & 56.3 & 52.6 & 33.5 \\
				&&Realized mean excess return&9.81& -1.10 & -0.93 & -0.98 & 2.34 \\
				&&Realized Sortino ratio&0.71& -0.08 & -0.07 & -0.07 & 0.18 \\
				&&Equal mean excess return p-value&& 0.003 & 0.002 & 0.001 & 0.001 \\
				\bottomrule
			\end{tabular}
		\end{threeparttable}
	}
	\floatfoot{\emph{Notes:} The table reports characteristics of the 215 portfolios selected with LP. State probabilities are specified as described in Section \ref{sec:phys}, but using the time-varying market risk premia plotted in Figure \ref{fig:mrptv}. Option contracts presenting a pure arbitrage opportunity are excluded. The top half of the table may be compared to the top quarter of Table \ref{table:mainresults}, while the bottom half may be compared to the top quarter of Table \ref{table:wideresults}.}
\end{table*}

We have found that if Equation \eqref{eq:BTZ09} is re-estimated using data from our sample period 2004--2021 then the estimated slope coefficient is statistically insignificant. (We found evidence of a significant relationship at a quarterly horizon, but not at the monthly horizon.) The time-variation in the estimated market risk premium plotted in Figure \ref{fig:mrptv} may therefore be pure noise. We believe, however, that the more pertinent lesson to be drawn from this example is that predictors of the market risk premium (or of state probabilities more generally) with a low signal-to-noise ratio should be avoided, even in cases where the signal is statistically significant.

\subsection{Time-varying asymmetry}

To explore the possibility that the degree of asymmetry of the distribution of normalized SPX returns varies systematically over our sample period we implemented a time-varying parametrization of the fitted SGT distribution. As discussed in Section \ref{sec:phys}, the SGT distribution is characterized by three parameters: shape ($k$), degrees of freedom ($\nu$) and asymmetry ($\lambda$). We used the implicit score-driven filter introduced in \citet{LangeOsDijk24} to obtain time-varying estimates of $\lambda$ while retaining the time-invariant estimates of $k$ and $\nu$ used in previous sections. The filter is implemented by setting $\hat{\lambda}_t$ equal to the full sample maximum likelihood estimate $\hat{\lambda}=-0.53$ on the first trading day in 2004, then iteratively choosing $\hat{\lambda}_t$ in subsequent trading days to maximize the penalized likelihood of the observed normalized SPX return on the day in question, with the penalty proportional to $(\hat{\lambda}_t-\hat{\lambda}_{t-1})^2$. See \citet{LangeOsDijk24} for a detailed description of the method.
\begin{figure*}[t!]
	\caption{Estimated time-varying asymmetry}\label{fig:asymmetrytv}
	\centering
	\subcaptionbox{Probability of positive normalized return}{
		\centering
		\begin{tikzpicture}[scale=3.5]
			\draw[black, thick,->] (0,0) -- (0,1.1);
			\draw[black, thick] (0,0) -- (1.3,0);
			\draw (0,0.3pt) -- (0,-0.8pt);
			\node[anchor=north,font=\scriptsize] at (0,-0.8pt) {$2004$};
			\draw ({(4*365+1)*1.3/6575},0.3pt) -- ({(4*365+1)*1.3/6575},-0.8pt);
			\node[anchor=north,font=\scriptsize] at ({(4*365+1)*1.3/6575},-0.8pt) {$2008$};
			\draw ({(8*365+2)*1.3/6575},0.3pt) -- ({(8*365+2)*1.3/6575},-0.8pt);
			\node[anchor=north,font=\scriptsize] at ({(8*365+2)*1.3/6575},-0.8pt) {$2012$};
			\draw ({(12*365+3)*1.3/6575},0.3pt) -- ({(12*365+3)*1.3/6575},-0.8pt);
			\node[anchor=north,font=\scriptsize] at ({(12*365+3)*1.3/6575},-0.8pt) {$2016$};
			\draw ({(16*365+4)*1.3/6575},0.3pt) -- ({(16*365+4)*1.3/6575},-0.8pt);
			\node[anchor=north,font=\scriptsize] at ({(16*365+4)*1.3/6575},-0.8pt) {$2020$};
			\draw (0.3pt,0) -- (-0.8pt,0);
			\draw (0.3pt,{1/3}) -- (-0.8pt,{1/3});
			\draw (0.3pt,{2/3}) -- (-0.8pt,{2/3});
			\draw (0.3pt,1) -- (-0.8pt,1);
			\node[left,font=\scriptsize] at (0,0) {$0.5$};
			\node[left,font=\scriptsize] at (0,{1/3}) {$0.55$};
			\node[left,font=\scriptsize] at (0,{2/3}) {$0.6$};
			\node[left,font=\scriptsize] at (0,1) {$0.65$};
			\draw[blue,thick] plot file {asymtv.txt};
			\draw[dashed] (0,{(.59-.5)/.15}) -- (1.3,{(.59-.5)/.15});
			\node[font=\footnotesize] at (.6,-.225) {Year};
			\node[rotate=90,font=\footnotesize] at (-.275,.5) {Probability};
		\end{tikzpicture}
		\hspace{.8cm}
	}
	\subcaptionbox{SGT distributions at asymmetry quintiles}{
		\centering
		\begin{tikzpicture}[scale=3.5]
			\draw[black, thick,->] (0,0) -- (0,1.1);
			\draw[black, thick] (0,0) -- (1.5,0);
			\draw (0,0.3pt) -- (0,-1pt)
			node[anchor=north,font=\scriptsize] {$-3$};
			\draw (0.3,0.3pt) -- (0.3,-1pt)
			node[anchor=north,font=\scriptsize] {$-2$};
			\draw (0.6,0.3pt) -- (0.6,-1pt)
			node[anchor=north,font=\scriptsize] {$-1$};
			\draw (0.9,0.3pt) -- (0.9,-1pt)
			node[anchor=north,font=\scriptsize] {$0$};
			\draw (1.2,0.3pt) -- (1.2,-1pt)
			node[anchor=north,font=\scriptsize] {$1$};
			\draw (1.5,0.3pt) -- (1.5,-1pt)
			node[anchor=north,font=\scriptsize] {$2$};
			\draw (0.3pt,0) -- (-0.8pt,0);
			\draw (0.3pt,{1/4}) -- (-0.8pt,{1/4});
			\draw (0.3pt,{2/4}) -- (-0.8pt,{2/4});
			\draw (0.3pt,{3/4}) -- (-0.8pt,{3/4});
			\draw (0.3pt,{4/4}) -- (-0.8pt,{4/4});
			\node[left,font=\scriptsize] at (0,0) {$0$};
			\node[left,font=\scriptsize] at (0,{1/4}) {$.05$};
			\node[left,font=\scriptsize] at (0,{2/4}) {$.1$};
			\node[left,font=\scriptsize] at (0,{3/4}) {$.15$};
			\node[left,font=\scriptsize] at (0,{4/4}) {$.2$};
			\draw[fill=blue!15] (0,0) rectangle (.1,{0.0067/.2});
			\draw[fill=blue!15] (.1,0) rectangle (.2,{0.0113/.2});
			\draw[fill=blue!15] (.2,0) rectangle (.3,{0.0173/.2});
			\draw[fill=blue!15] (.3,0) rectangle (.4,{0.0242/.2});
			\draw[fill=blue!15] (.4,0) rectangle (.5,{0.0327/.2});
			\draw[fill=blue!15] (.5,0) rectangle (.6,{0.0540/.2});
			\draw[fill=blue!15] (.6,0) rectangle (.7,{0.0614/.2});
			\draw[fill=blue!15] (.7,0) rectangle (.8,{0.0778/.2});
			\draw[fill=blue!15] (.8,0) rectangle (.9,{0.1136/.2});
			\draw[fill=blue!15] (.9,0) rectangle (1,{0.1470/.2});
			\draw[fill=blue!15] (1,0) rectangle (1.1,{0.1681/.2});
			\draw[fill=blue!15] (1.1,0) rectangle (1.2,{0.1476/.2});
			\draw[fill=blue!15] (1.2,0) rectangle (1.3,{0.0845/.2});
			\draw[fill=blue!15] (1.3,0) rectangle (1.4,{0.0278/.2});
			\draw[fill=blue!15] (1.4,0) rectangle (1.5,{0.0080/.2});
			\draw[cyan,thick] plot file {sgtpdf1.txt};
			\draw[magenta,thick] plot file {sgtpdf2.txt};
			\draw[olive,thick] plot file {sgtpdf3.txt};
			\draw[teal,thick] plot file {sgtpdf4.txt};
			\node[font=\footnotesize] at (.75,-.225) {Normalized SPX return};
			\node[rotate=90,font=\footnotesize] at (-.225,.5) {Probability};
		\end{tikzpicture}
		\hspace{.7cm}
	}
	\floatfoot{\emph{Notes:} Panel (a) plots the probability of a positive normalized SPX return implied by our time-varying parametrization of the SGT distribution. Panel (b) displays the SGT distributions corresponding to the first, second, third and fourth quintiles of the time-varying asymmetry parameter. These are overlaid on a histogram of normalized SPX returns, shown also in panel (a) of Figure \ref{fig:PDFs}.}
\end{figure*}

\begin{table*}[b!]
	\centering
	{\fontsize{9}{10.8}\selectfont
		\begin{threeparttable}
			\caption{Results with time-varying asymmetry}\label{table:asymmetrytv}
			\begin{tabular}{clccccc}
				\toprule
				&&&\multicolumn{4}{c}{Enhanced portfolio}\\\cmidrule{4-7}
				Strike range&& SPX & $S=1$ & $S=10$ & $S=100$ & $S=1000$\\
				\midrule
				\multirow{4}{*}{Moderate}&Pctg.\ premia $>0.1\% $ mkt.\ investment&&  7.4 &  6.0 &  4.2 &  2.8 \\
				&Realized mean excess return&9.81& 8.73 & 8.46 & 8.26 & 8.91 \\
				&Realized Sortino ratio&0.71& 0.64 & 0.62 & 0.60 & 0.65 \\
				&Equal mean excess return p-value&& 0.145 & 0.063 & 0.033 & 0.085 \\
				\midrule
				\multirow{4}{*}{Wide}&Pctg.\ premia $>0.1\% $ mkt.\ investment&& 58.1 & 51.6 & 37.2 & 15.3 \\
				&Realized mean excess return&9.81& 7.26 & 5.51 & 5.82 & 6.81 \\
				&Realized Sortino ratio&0.71& 0.48 & 0.39 & 0.42 & 0.50 \\
				&Equal mean excess return p-value&& 0.502 & 0.179 & 0.100 & 0.050 \\
				\bottomrule
			\end{tabular}
		\end{threeparttable}
	}
	\floatfoot{\emph{Notes:} The table reports characteristics of the 215 portfolios selected with LP. State probabilities are specified as described in Section \ref{sec:phys}, but using the time-varying asymmetry parameter plotted in Figure \ref{fig:asymmetrytv}. Option contracts presenting a pure arbitrage opportunity are excluded. The top half of the table may be compared to the top quarter of Table \ref{table:mainresults}, while the bottom half may be compared to the top quarter of Table \ref{table:wideresults}.}
\end{table*}

Panel (a) of Figure \ref{fig:asymmetrytv} plots the estimates of time-varying asymmetry we obtained for each of the 215 dates at which option portfolios are formed. To ease interpretation we convert each estimated value of $\lambda$ to the implied probability of a positive normalized SPX return. We see that this probability varies from roughly 0.5 to 0.62, compared to a probability of approximately 0.59 using the time-invariant maximum likelihood estimate $\hat{\lambda}=-0.53$. In panel (b) of Figure \ref{fig:asymmetrytv} we plot the SGT distributions corresponding to the first, second, third and fourth quintiles of the time-varying estimates of $\lambda$, overlaying these on a histogram of normalized SPX returns.

Key characteristics of the option portfolios selected using the SGT parametrization with time-varying asymmetry are reported in Table \ref{table:asymmetrytv}. Portfolio performance is qualitatively similar, but somewhat poorer, than was the case using the time-invariant parametrization. This suggests that our time-varying estimates of asymmetry may be excessively noisy.

The filtering procedure of \citet{LangeOsDijk24} uses a tuning parameter to control the smoothness of parameter variation over time. We have found, after repeating our analysis with a wide range of tuning parameter values, that the performance of our selected option portfolios becomes worse as smoothness is reduced. We observe the best performance---which is still no better than that of the market portfolio---when asymmetry is specified to be time-invariant. Time-variation in asymmetry may well be present, but our understanding of it is insufficient to usefully inform portfolio selection.


\end{document}